\documentclass[10pt,journal,compsoc]{IEEEtran}

\ifCLASSOPTIONcompsoc
  \usepackage[nocompress]{cite}
\else
  \usepackage{cite}
\fi

\ifCLASSINFOpdf
\else
\fi

\usepackage{graphicx}
\usepackage{soul}
\usepackage{subfig}
\usepackage[fleqn]{amsmath}
\usepackage{relsize}
\usepackage{algorithm}
\usepackage[noend]{algpseudocode}
\usepackage{varwidth}
\usepackage{ragged2e}
\usepackage{booktabs} 
\usepackage[percent]{overpic}
\usepackage{multirow}
\usepackage{bm}
\usepackage{enumitem}
\usepackage[normalem]{ulem}
\usepackage{tikz}

\soulregister\cite7
\soulregister\ref7
\soulregister\pageref7

\usepackage[algo2e]{algorithm2e} 

\usepackage[super]{nth}
\usepackage[numbers,sort&compress]{natbib}

\begin{document}
%
\title{CoPA: \underline{Co}ld \underline{P}age \underline{A}wakening to Overcome Retention Failures in STT-MRAM Based I/O Buffers}
%
%
%
%

\newcommand{\textoverline}[1]{$\overline{\mbox{#1}}$}
\newcommand{\specialcell}[2][c]{%
	\begin{tabular}[#1]{@{}l@{}}#2\end{tabular}}

\author{Mostafa Hadizadeh\textsuperscript{1}\thanks{1. Mostafa Hadizadeh, Elham Cheshmikhani, Maysam Rahmanpour, and Hossein Asadi (corresponding author) are with the Department of Computer Engineering, Sharif University of Technology, Tehran 11155-11365, Iran. Emails: mhadizadeh@ce.sharif.edu, elham.cheshmikhani@sharif.edu, rahmanpour@ce.sharif.edu, asadi@sharif.edu}, Elham Cheshmikhani\textsuperscript{1}, Maysam Rahmanpour\textsuperscript{1}, \\Onur Mutlu\textsuperscript{2}\thanks{2. Onur Mutlu is with the Department of Computer Science, ETH Zurich, Switzerland. Email: omutlu@gmail.com}, and Hossein Asadi\textsuperscript{1}
}

\IEEEtitleabstractindextext{%
	{\justify
\begin{abstract}
	Performance and reliability are two prominent factors in the design of data storage systems. To achieve higher performance, recently storage system designers use $Dynamic$ $RAM$ (DRAM)-based buffers. The volatility of DRAM brings up the possibility of data loss and data inconsistency. Thus, a part of the main storage is conventionally used as the journal area to be able of recovering unflushed data pages in the case of power failure. Moreover, periodically flushing buffered data pages to the main storage is a common mechanism to preserve a high level of reliability. This scheme, however, leads to a considerable increase in storage write traffic, which adversely affects the performance. To address this shortcoming, recent studies offer a small $Non$$-$$Volatile$ $Memory$ (NVM) as the $Persistent$ $Journal$ $Area$ (PJA) along with DRAM as an efficient approach to overcome DRAM vulnerability against power failure while effectively reducing storage write traffic. This approach, named $NVM$$-$$Backed$ $Buffer$ (NVB-Buffer), features from advantages of NVMs and addresses DRAM shortcomings. In this paper, we employ the most promising technologies for PJA among the emerging technologies, which is $Spin$$-$$Transfer$ $Torque$ $Magnetic$ $Random$ $Access$ $Memory$ (STT-MRAM) to meet the requirements of efficient PJA by providing high endurance, non-volatility, and DRAM-like latency. Despite these advantages, STT-MRAM faces major reliability challenges, i.e. \textit{Retention Failure}, \textit{Read Disturbance}, and \textit{Write Failure}, which have \textit{not} been addressed in previously suggested NVB-Buffers. In this paper, we first demonstrate that the retention failure is the dominant source of errors in NVB-Buffers as it suffers from long and unpredictable page idle intervals (i.e., the time interval between two consecutive accesses to a PJA page). Then, we propose a novel NVB-Buffer management scheme, named, $\underline{Co}ld$ $\underline{P}age$ $\underline{A}wakening$ (CoPA), which predictably reduces the idle time of PJA pages. To this aim, CoPA employs $Distant$ $Refreshing$ to periodically overwrite the vulnerable PJA page contents by opportunistically using their replica in DRAM-based buffer. 
We compare CoPA with the state-of-the-art schemes over several well-known storage workloads based on physical journaling. Our evaluations show that CoPA significantly reduces the maximum page idle time, which leads to three orders of magnitude lower failure rate with negligible performance degradation (1.1\%) and memory overhead (1.2\%).	
\end{abstract}}

}

\maketitle

\IEEEdisplaynontitleabstractindextext

%
\IEEEpeerreviewmaketitle
\markboth{IEEE Transactions on Parallel and Distributed Systems,~Vol.~XX, No.~X, March~XXXX}%
{CoPA: Cold Page Awakening to Overcome Retention Failures in STT-MRAM Based I/O Buffers}

	\section{Introduction}

Storage systems use \textit{Dynamic Random Access Memory} (DRAM)-based buffers to mitigate the considerable latency gap between storage and main memory \cite{Hibachi,DF-LRW,TICA,H-ARC}. 
DRAM benefits from low access time, high density, and unlimited lifetime, which makes it a suitable candidate for these applications. However, data loss due to power failure is the major drawback of DRAM-based buffers because of its volatility. Thus, DRAM-based buffers exploit mechanisms such as periodic flush \cite{DF-LRW,ATC05} and/or partial use of main storage as the journal area \cite{PhysicalJournaling,SN,NV-Log,Zhang-ICCD,Unified-NVM-DRAM} to recover data in the case of power failure. 
Nevertheless, these techniques significantly increase storage write traffic.


Thanks to emerging \textit{Non-Volatile Memories} (NVMs) \cite{TPDS1,IEEETR3,IEEETR1,EC-TR,TWOLRU,STT-MainMem,STRATA,MUTLU-ThyNVM, IEEETR2}, a recent approach to design an efficient buffer is to use a relatively small NVM as the journal area \cite{DF-LRW,NV-Log} along with DRAM-based buffer. This approach is referred to as \textit{NVM-Backed Buffer} (NVB-Buffer) in this study. In this approach, the buffer takes advantage of NVM persistency to reliably store dirty data pages,  beside low latency of DRAM for fast accesses. To this end, NVM is recognized as $Persistent$ $Journal$ $Area$ (PJA) where a copy of all dirty data pages is stored in NVM. In this case, there always exists a \textit{persistent} valid copy of the DRAM data page. Thus, NVB-Buffer can use PJA data pages for data recovery in the case of power failure while considerably reducing storage write traffic \cite{DF-LRW}.   

\begin{figure}[t]
	\includegraphics[width=9cm]{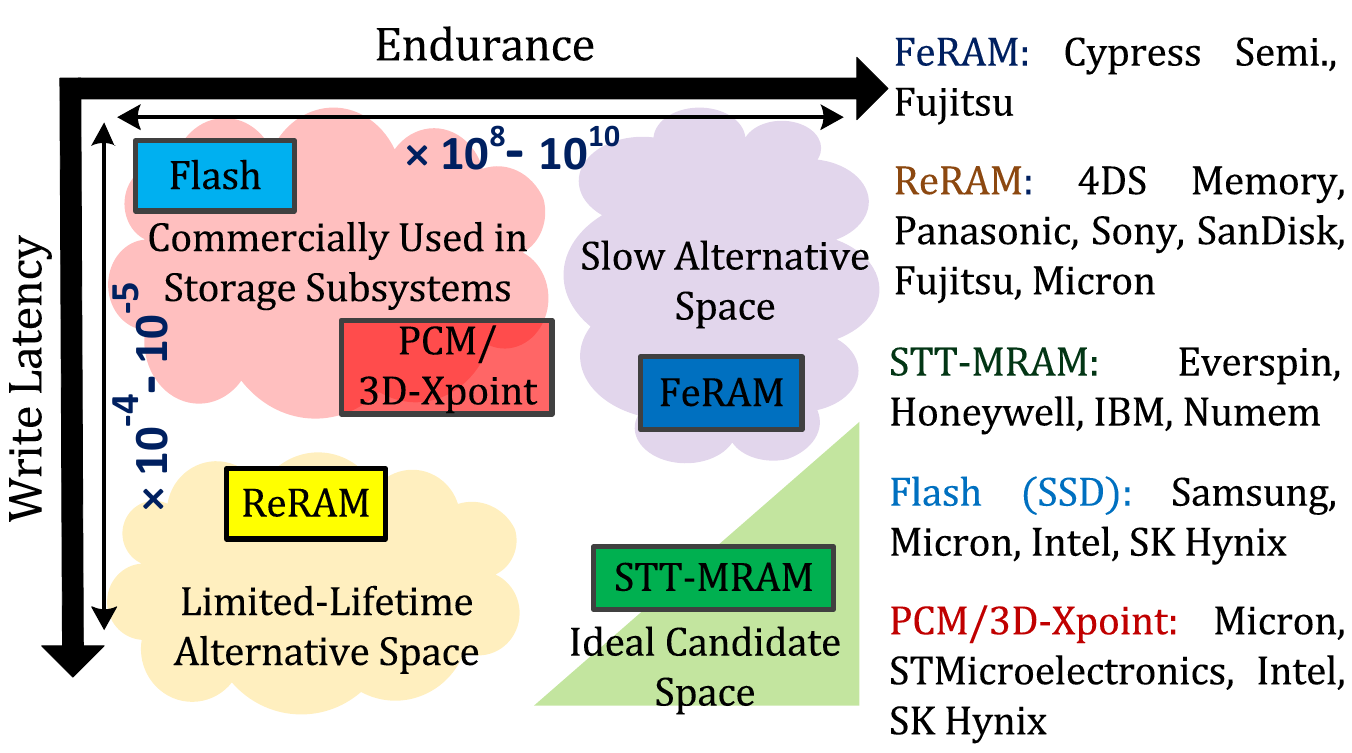}
	\centering\caption{Overview of existing NVMs with respect to ideal PJA requirements \cite{TA-LRW,Endurance-STT,Tarihi,STT-MainMem,NVMSurvey,NVMFS,3dxpoint,FRAM-Cypress,rramsite,mramsite,flashproducers,PCM-site,FRAMFujitsu,STMElectronics,RRAMPaper}}
	\label{Tech-Candidates}
\end{figure}

Previous studies attempt to use \textit{Phase Change Memory} (PCM) technology in NVB-Buffer \cite{Unified-NVM-DRAM,PCM1,PCM2}. PCM, however, suffers from limited endurance, poor write performance, and considerable write power \cite{Ali-TETC,DATESalkhordeh,MUTLU-CASES,MUTLU-TACO,PCM0}, which makes it an ineffective candidate for NVB-Buffers. 
Moreover, existing NVM candidates such as Flash, Ferroelectric RAM (FeRAM), 3D-Xpoint, and Resistive RAM (ReRAM) either suffer from high write latency (3D-XPoint, ReRAM, and Flash) or limited lifetime (3D-XPoint, Flash, and FeRAM) \cite{Tarihi,STT-MainMem,NVMSurvey,NVMFS,3dxpoint}. Fig. \ref{Tech-Candidates} illustrates the existing memory technologies with respect to PJA requirements. 
Among these candidates, \textit{Spin-Transfer Torque Magnetic Random Access Memory} (STT-MRAM) technology meets the PJA requirements of non-volatility \cite{STT-MainMem}, high endurance \cite{Endurance-STT}, and DRAM-comparable performance \cite{STT-MainMem}, as well as 
high density and negligible leakage current \cite{TA-LRW}. Therefore, compared to the other PJA candidates, STT-MRAM is considered as one of the most promising technologies for NVB-Buffers \cite{Endurance-STT,Tarihi,STT-MainMem,IEEETR1,NVMFS,NVMSurvey}.

Despite STT-MRAM advantages, it faces three reliability challenges such as \textit{Retention Failure}, \textit{Write Failure}, and \textit{Read Disturbance}. Bit errors in a PJA page due to these challenges lead to data loss in the case of power failure, as there is no valid copy of the page in the main storage. Moreover, previous studies mainly aim to enhance performance or reduce journaling overhead \cite{DF-LRW,Unified-NVM-DRAM,SCHEN-TCAD18,SN}. To our knowledge, \textit{none} of the prior studies considers the reliability of STT-MRAM based NVB-Buffers, while the previous schemes to overcome STT-MRAM retention failure are either expensive or not applicable as they increase the probability of read disturbance or cause performance degradation \cite{Sanitizer,RetentionHPCA11,SeleRef,FlowPaP}. 

In this paper, we first investigate existing NVB-Buffer schemes using several well-known storage workloads to explore their vulnerability to STT-MRAM error sources. Our extensive set of experiments on \textit{Microsoft Research Cambridge} (MSRC) traces \cite{MSRC} show that PJA pages in NVB-Buffer suffer from long idle time, varying from 34.5 to 100.8 minutes, which makes them highly vulnerable to retention failure. Moreover, technology downscaling makes retention failure more severe \cite{ITJ,TA-LRW}. Based on our experiments, retention failure is the dominant source of PJA failures in recent technology nodes, as the probability of data loss due to retention failure is five orders of magnitude higher than write failure, on average \cite{ITJ,TA-LRW}. However, read disturbance is not a serious challenge in PJA as the PJA pages are inherently used for data residency (pages are just read in the case of data recovery due to power failure). 

We then propose a novel management scheme for NVB-Buffers, named, \textit{\underline{Co}ld \underline{P}age \underline{A}wakening} (CoPA) to efficiently reduce the idle time of PJA pages. CoPA reduces idle intervals using \textit{Distant Refreshing}, where it overwrites PJA pages using error-free replicas in DRAM-based buffers.
CoPA also differentiates PJA pages based on their vulnerabilities to retention failure and prevents recently written pages from refreshing. To show the effectiveness of Distant Refreshing, we examine and compare the proposed scheme against the conventional refreshing. Our experiments show that Distant Refreshing reduces the failure rate by 89.9\%, on average, compared to conventional refreshing. Conventional refreshing employs read-correct-write approach, which increases the probability of read disturbance and, consequently, leads to increase in total failure rate.

CoPA aims at providing different levels of reliability and can be tuned depending on the application requirements. Our experimental evaluations based on the physical journaling \cite{ATC05,PhysicalJournaling} are performed across twelve storage workloads from MSRC \cite{MSRC}. CoPA is tuned to guarantee an upper bound for maximum idle time of PJA pages in such a way that reduces the maximum idle time of PJA pages by an average of 53.5$\times$ (up to 66.9$\times$) compared to the state-of-the-art NVB-Buffer management schemes \cite{DF-LRW,Unified-NVM-DRAM,Zhang-ICCD}. CoPA results in three orders of magnitude failure rate reduction, with negligible performance (an average of 1.1\%) and memory (1.2\%) overhead. We also compare CoPA with the conventional periodic flush-enabled scheme, which provides high reliability at the cost of considerable storage write traffic. Tuning CoPA in favor of reliability leads to an average of 43\% (up to 81.9\%) response time reduction.

The main \textbf{contributions} of this paper are as follows:
\begin{itemize}
	\item This is the \textit{first} study that investigates the reliability of STT-MRAM based NVB-Buffers. We examine the state-of-the-art schemes and show that retention failure is the main contributor to PJA errors. PJA pages suffer from long page idle times, which significantly degrades NVB-Buffer reliability due to high probability of retention failure.
	\item We propose a novel NVB-Buffer management scheme called CoPA, aims at reducing the idle time of PJA pages to mitigate their vulnerability to retention failure. It overwrites PJA pages by employing their valid replica in DRAM while categorizes PJA pages based on their vulnerability to retention failure and prevents from overwriting of recently-written pages. 
	\item Our evaluations show that CoPA can be tuned according to the application reliability requirements, while considerably improves performance compared to the conventional reliable schemes. 
	\item We extensively evaluate CoPA in terms of failure rate and reliability. The results illustrate that CoPA significantly reduces the error rate by three orders of magnitude with very negligible performance and memory overhead compared to the state-of-the-art schemes (1.1\% and 1.2\% on average, respectively). 
\end{itemize}
\section{Background}
\subsection{NVM-Backed Buffer}
\begin{figure}[t]
	\includegraphics[width=9.5cm]{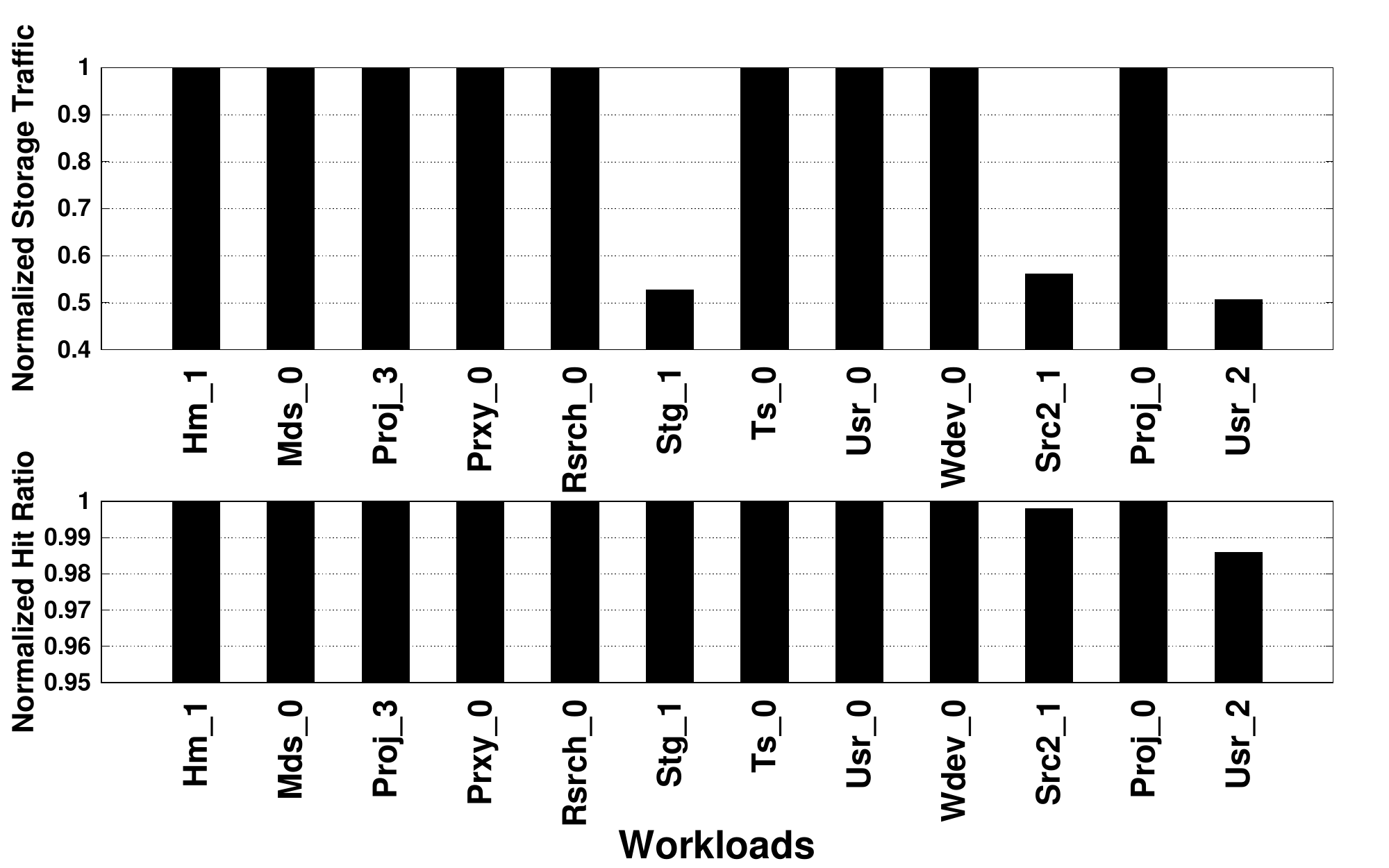}
	\centering\caption{Hit ratio and storage traffic of NVB-Buffer normalized to Hyb-Buffer}
	\label{HybBuffer}
\end{figure}

Buffering I/O requests is one of the conventional solutions for performance improvement due to reducing storage write traffic and buffer hits in storage systems \cite{TICA}. In addition to performance improvement, in high-performance \textit{Solid-State Drive} (SSD)-equipped storage systems, reducing storage write traffic leads to the lifetime improvement, as Flash-based SSDs suffer from limited endurance \cite{ECI,MUTLU-DSN,DEFT}. DRAM-based buffers, however, face with unrecoverable data loss in case of power failure due to volatile characteristic of DRAM. To address this challenge, several techniques such as a) periodically flushing buffered pages to the main storage \cite{DF-LRW,ATC05} and b) partial use of the main storage as the journal area \cite{PhysicalJournaling,SN,NV-Log,Zhang-ICCD,Unified-NVM-DRAM} are exploited for data recovery in the case of power failure. Nevertheless, these techniques lead to a considerable increase in storage write traffic \cite{DF-LRW}.

Taking advantage of NVMs persistency along with DRAM performance is one of the promising schemes for designing efficient buffers \cite{Hibachi,DF-LRW}. \textit{NVM-Backed Buffer} (NVB-Buffer) is one of the recent schemes that provides efficient I/O buffers. In this scheme, DRAM-based buffer is equipped with a relatively small NVM as a \textit{Persistent Journal Area} (PJA), which \textit{only} tracks the dirty data pages of DRAM. In NVB-Buffers, the NVM pages are only read in the case of data recovery while in other scenarios, the accesses to NVM are only writing dirty pages. NVM pages are flushed whether NVM becomes full or their corresponding pages in volatile buffer are evicted. Thus, DRAM pages always have a persistent copy that prevents data loss in case of power failure.

We analyze the efficiency of NVB-Buffer compared to the conventional \textit{One-tier Hybrid Buffer} (Hyb-Buffer) to understand the impact of adding NVM space to operational buffer capacity on system performance. In Hyb-Buffers, DRAM and NVM are managed as a single tier, where the dirty data pages are redirected to both memories to prevent data loss in the case of power failure. Also, evicted data pages from DRAM are admitted to the NVM while NVM hit leads to migration of the page to the DRAM (if the page is dirty, NVM still buffers its copy). Fig. \ref{HybBuffer} shows the hit ratio and storage traffic of NVB-Buffer normalized to the Hyb-Buffer. The evaluations are performed based on a buffer consists of 8GB DRAM and 512MB NVM. Although both schemes provide the same performance behavior in workloads with working set smaller than the buffer capacity, NVB-Buffer significantly reduces storage traffic in workloads with working set greater than the buffer capacity. NVB-Buffer alleviates storage traffic by 47.4\%, 43.9\%, and 49.4\% in Stg\_1, Src2\_1, and Usr\_2, respectively. Hyb-Buffer, however, provides slightly higher hit ratio in the same set of workloads (up to 1.4\%). 

The traffic reduction is achieved by providing more presence time for dirty data pages in the buffer. Inserting the evicted data pages from DRAM to the NVM increases the number of NVM evictions. If the evicted page from NVM is dirty, as a DRAM dirty page should have a persistent copy, the page is written to the main storage and the status of its corresponding page in DRAM is changed to clean. Thus, even if the page still resides in the buffer, it is already flushed to the main storage. We also evaluate the scenario where NVM tracks dirty pages even after the eviction of their corresponding copy from DRAM to provide less storage traffic and more chance for buffer hit. The improvement of this scenario, however, is less than 0.01\%, at the best case.
\begin{figure}[t]
	\includegraphics[width=8cm]{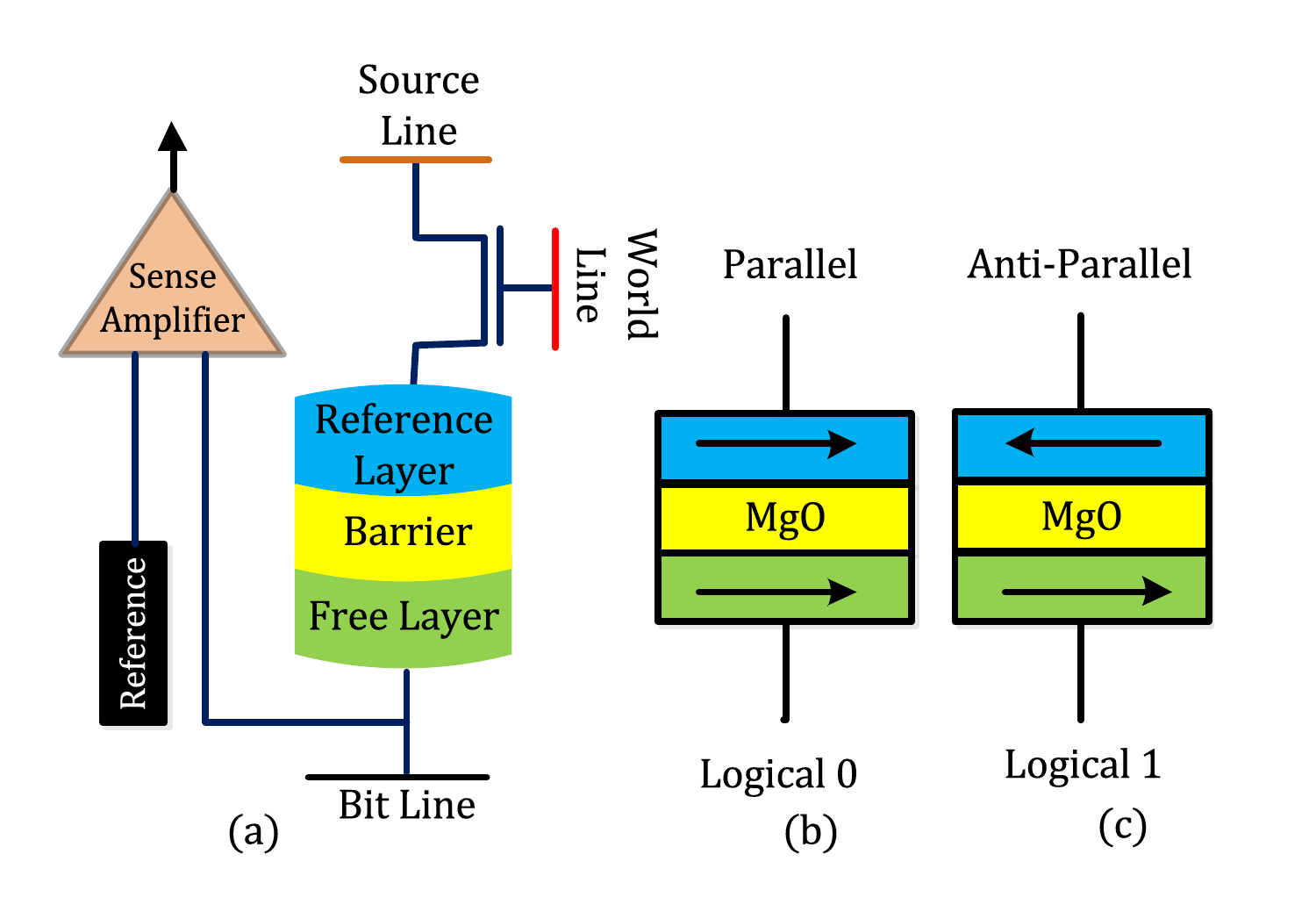}
	\centering\caption{a) STT-MRAM cell structure, b) parallel state, and c) anti-parallel state \cite{STT-MainMem,EDCC-Cheshmikhani}}
	\label{STT-Basics}
\end{figure}

\subsection{STT-MRAM Basics}
The most promising technology among emerging NVMs is STT-MRAM, which can be used in NVB-Buffers. An STT-MRAM cell consists of an access transistor and a \textit{Magnetic Tunnel Junction} (MTJ), which is responsible for data storage, shown in Fig. \ref{STT-Basics}-a. MTJ is made of an insulating layer (Magnesium Oxide), and two magnetic layers named \textit{Free Layer} and \textit{Reference Layer}. The magnetic field direction of the reference layer is fixed while the direction of the free layer can be changed and determine the constructed resistance value. If the free layer is paralleled with the reference layer in the case of flowing write current $(I_{write})$ from the free layer to the reference layer, MTJ will be in low resistance state, which is the logical `0' value (Fig. \ref{STT-Basics}-b). An anti-parallel pattern between the spin directions of the free layer and the reference layer is due to $I_{write}$ flowing from the reference layer to the free layer, which results in high MTJ resistance and logical `1' value (Fig. \ref{STT-Basics}-c) \cite{STT-MainMem,EDCC-Cheshmikhani}.



STT-MRAM provides promising features in terms of endurance, performance, power consumption, and scalability. It is expected that STT-MRAM partly replaces technologies such as NOR Flash, \textit{Static RAM} (SRAM), and DRAM, while it is estimated that standalone \textit{Magnetic RAM} (MRAM) and STT-MRAM baseline revenues reach \$3.8B in 2029 \cite{STT-Revnue}. However, STT-MRAM is threatened by three types of failures. To read the stored data in an STT-MRAM cell, a small current is applied to the free layer to recognize \textit{high} or \textit{low} resistance level. It is probable that this small current leads to a bit flip in the cell, which is called \textit{Read Disturbance}. To write data, a write current is applied to change the magnetic field direction of the free layer. Nevertheless, the cell content may not change during the write pulse, which leads to a \textit{Write Failure}. Moreover, stochastic characteristics of STT-MRAM leads to inadvertent bit flip, as the stored data would be changed, without applying any current to the MTJ, and leads to a \textit{Retention Failure}. To make STT-MRAM a widely commercialized replacement for conventional technologies to be used in NVB-Buffers, it is a must to address its reliability challenges.

\begin{figure}[t]
	\includegraphics[width=9cm]{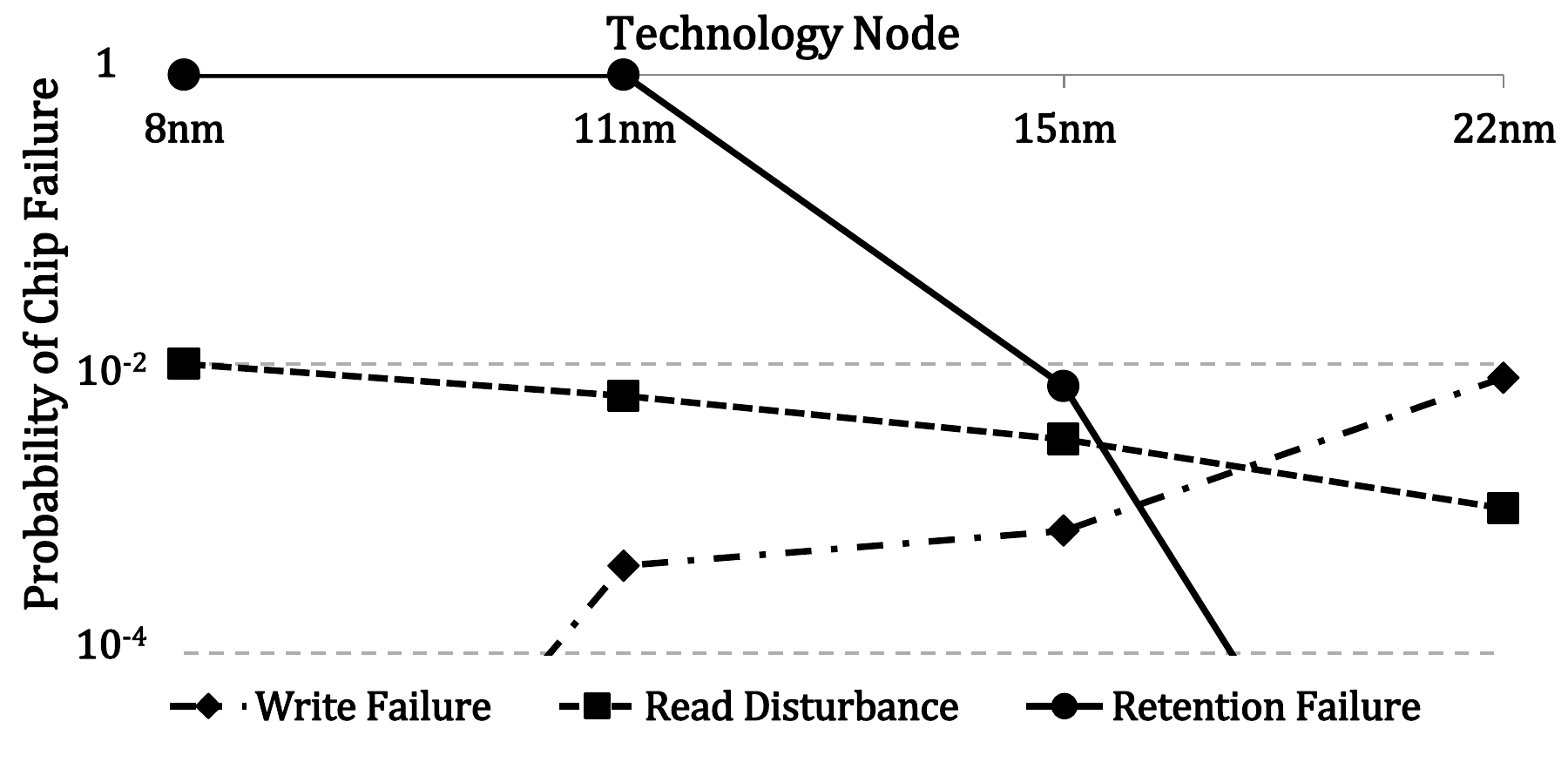}
	\centering\caption{Footprint of each error type on the probability of STT-MRAM failure \cite{ITJ,ProcessVariation}}
	\label{Node-Downscaling}
\end{figure}


\section{Motivation}
\subsection{Reliability Challenges of STT-MRAM Based PJA}
Any candidate technology to be used in PJA needs to provide three features, a) \textbf{non-volatility}: PJA must be capable of retaining data even in the case of power failure to be used for data recovery, b) \textbf{high endurance}: due to extreme write pressure on PJA and its relatively small size (compared to DRAM), it should be able to endure a high number of writes without lifetime degradation, and c) \textbf{DRAM-comparable write latency}: to prevent inconsistency in reading data from PJA and the buffer (because of large difference between their write performance), PJA delay should be less or equal to DRAM buffers. 


Among emerging technologies, STT-MRAM is the most suitable for PJA. PCM, 3D-Xpoint, and Flash neither can provide PJA endurance nor its performance requirements \cite{Tarihi,NVMSurvey,NVMFS,3dxpoint}. Although FeRAM offers high endurance, its write latency is considerably higher than STT-MRAM \cite{STT-MainMem,NVMSurvey,NVMFS}. ReRAM is another possible PJA candidate, which provides promising performance feature. ReRAM-based PJA suffers from limited lifetime \cite{STT-MainMem,NVMSurvey,NVMFS,Endurance-STT}. Therefore, due to STT-MRAM high endurance and low write latency (compared to FeRAM, PCM, Flash, and 3D-Xpoint), it is the most promising PJA candidate. On the other hand, STT-MRAM provides lower write energy compared to the mentioned technologies \cite{NVMFS}. However, to provide an efficient NVB-Buffer, the aforementioned reliability challenges of STT-MRAM should be carefully addressed.
Fig. \ref{Node-Downscaling} shows the footprint of each error type on the total failure rate of a 32MB STT-MRAM cache \cite{ITJ}. 
The probability of write failure is considerably reduced by technology downsizing, while the probability of read disturbance is increased. Although read disturbance is still a serious concern in processor cache levels and main memory \cite{ROBIN,DATECheshmikhani,Sanitizer}, from the PJA perspective, read disturbance is not the main challenge as PJA is inherently exploited for data residency. Thus, the stored data in PJA is only read for the sake of data recovery due to power failure or system crash (once during its lifetime). 
By technology node downsizing, however, retention failure becomes the dominant source of STT-MRAM errors \cite{ITJ,ProcessVariation,StatisticalSTTRetention,EC-TR} and leads to a significant reduction in STT-MRAM \textit{Mean Time To Failure} (MTTF), as shown in Fig. \ref{Node-Downscaling} \cite{TA-LRW}. Therefore, PJA data pages could only be valid for a very short time intervals. Increasing the probability of retention failure is one of the main challenges of employing STT-MRAM as PJA, since it threatens data retaining, which is essential for PJA. 
\subsection{Impact of NVB-Buffer Accesses on PJA Page Idle Time}
\begin{figure}[t]
	\includegraphics[width=8.5cm]{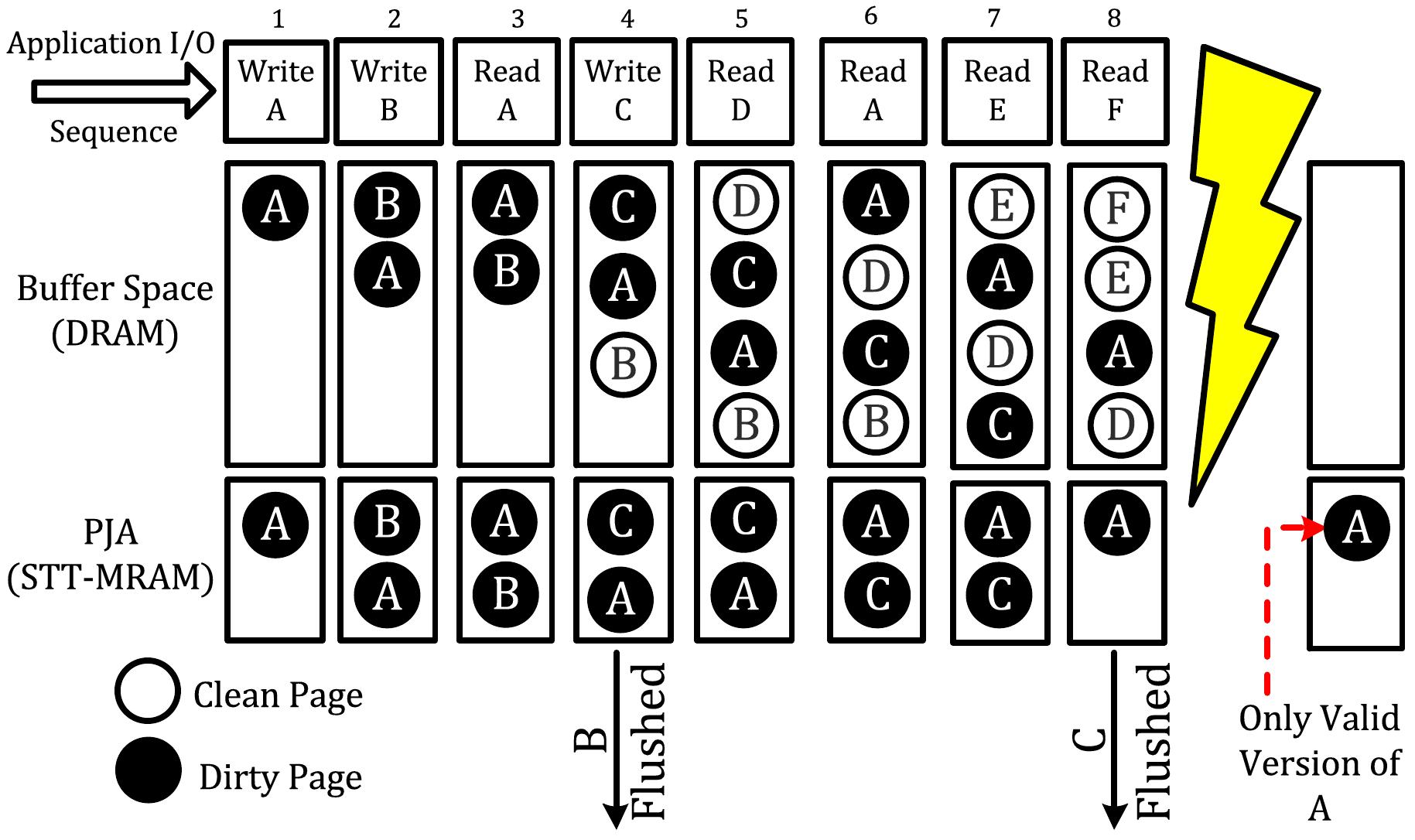}
	\centering\caption{An example of NVB-Buffer access pattern impact on PJA page idleness}
	\label{Example}
\end{figure}

To examine the idle time of PJA pages, which is the main contributor of retention failure rate, a motivational example of how NVB-Buffer access pattern affecting the PJA page idleness is provided as shown in Fig. \ref{Example}. It is assumed that the buffer and PJA capacity are four and two pages (for the sake of visibility), respectively, while both use \textit{Least Recently Used} (LRU) replacement policy and physical journaling \cite{PhysicalJournaling} is employed. \nth{1} and \nth{2} accesses lead to the insertion of \textit{A} and \textit{B} in the buffer and PJA (as both requests are write-requests). The \nth{3} access leads to reading \textit{A}, which calls to update both queues. However, \textit{A} is read from the buffer and the page \textit{A} in PJA is not either read or refreshed, so it is still idle in PJA. At \nth{4} access, \textit{C} is written, which leads to eviction of \textit{B} from PJA. Thus, \textit{B} is written to the main storage and interpreted as a clean page. 

From the \nth{5} access to the \nth{7} access, the buffer is filled, \textit{A} is redirected to the \textit{Most Recently Used} (MRU) position, and \textit{B} is evicted from NVB-Buffer. In the \nth{8} access, \textit{F} is inserted into the buffer leading to the eviction of \textit{C} from the buffer. As \textit{C} is a dirty page, it is redirected to the main storage and its corresponding page is also evicted from PJA. 
\textit{A} is admitted to the NVB-Buffer at the \nth{1} access and then is read multiple times, which prevents \textit{A} from redirecting to the main storage. In the case of system crash or power failure, the only dirty page that should be recovered is \textit{A}, which is exposed to retention failure due to its long idle time. 

This instance demonstrates the access sequences on a small scale. As it shows, a long idle time can be caused for a page after a while, during the access sequences. In this case, long durations should be shortened to decrease the retention failure rate. Therefore, a reliable NVB-Buffer management scheme should seriously consider mechanisms to alleviate the vulnerability of pages such as \textit{A}. 

\subsection{Impact of Various Error Types on PJA Failures}
We provide an illustration about the contribution of retention failure in PJA failures compared to other reliability challenges. We set up an experiment to investigate the impact of each error type on the failure of PJA pages. NVB-Buffer consists of 8GB DRAM and 512MB STT-MRAM based PJA, while employing physical journaling with 4KB page size.

\subsubsection{Formulation}
Retention failure depends on the idle intervals and the probability of its occurrence for a cell is according to (1) \cite{ITJ}:
{\normalsize\begin{equation}P_{rf_{cell}}(t)=1-e^\frac{-t}{e^\Delta},
	\end{equation}
}where $t$ and $\Delta$ represent idle time and thermal stability factor, respectively. Suppose that each STT-MRAM page consists of 512 64-bit word, and each word is protected by a \textit{Single Error Correction-Double Error Detection} (SEC-DED) code. The probability of data loss due to retention failure of a page for idle time equal to \textit{t} is according to (2):
{\normalsize 
	\begin{multline}
		P_{dl\_rf_{page}}(t)= 1 - [(1-P_{rf_{cell}}(t))^k + \\k\times(1-P_{rf_{cell}}(t))^{k-1}\times(P_{rf_{cell}}(t))]^W,
	\end{multline}
}where \textit{k} is the number of bits in a word and \textit{W} is the number of words in a page, which are 64 and 512, respectively.

The probability of data loss of a page for all idle intervals is according to (3):
{\normalsize
	\begin{multline}
		P_{dl\_rf_{all\_intervals}} = 1 - \prod_{i=1}^{n} (1-P_{dl\_rf_{page}}(t_i)),
	\end{multline}
}where \textit{n} is the number of intervals while $t_i$ represents the page idle time during interval \textit{i}.

Write failure is another error type that is likely to happen in PJA pages. The probability of write failure for an STT-MRAM cell is according to (4) :
{\normalsize 
	\begin{multline}
		P_{wf_{cell}}=e^\frac{-t_{write}\times2\times \mu_{\beta} \times p\times (I_{write}-I_{C_o})}{c + (e\times m\times (1+p^2))\times ln(\pi^2\times \Delta/4)},
	\end{multline}
}where $t_{write}$ represents the width of write pulse, $\mu_\beta$ is \textit{Bohr} magneton, \textit{p} is the tunneling spin polarization, and $I_{write}$ represents write pulse width, while \textit{c} and \textit{m} are \textit{Euler} constant and magnetic momentum of the free layer, respectively. 

The probability of data loss due to write failure for a SEC-DED protected page is according to (5):
{\normalsize 
	\begin{multline} 
		P_{dl\_wf_{page}}=1 - [[(1-P_{wf_{cell}})^k + \\ k\times(1-P_{wf_{cell}})^{k-1}\times(P_{wf_{cell}})]^W]^N,
	\end{multline} 
}where \textit{k} is the number of bits in a word, \textit{W} is the number of words in a cell, and \textit{N} represents the number of writes committed to a page. In NVB-Buffer, since PJA content is just read once for data recovery upon power failure, the impact of read disturbance on PJA failure is highly negligible.

\subsubsection{PJA Failure Characterization}
\begin{figure}[t]
	\includegraphics[width=9cm]{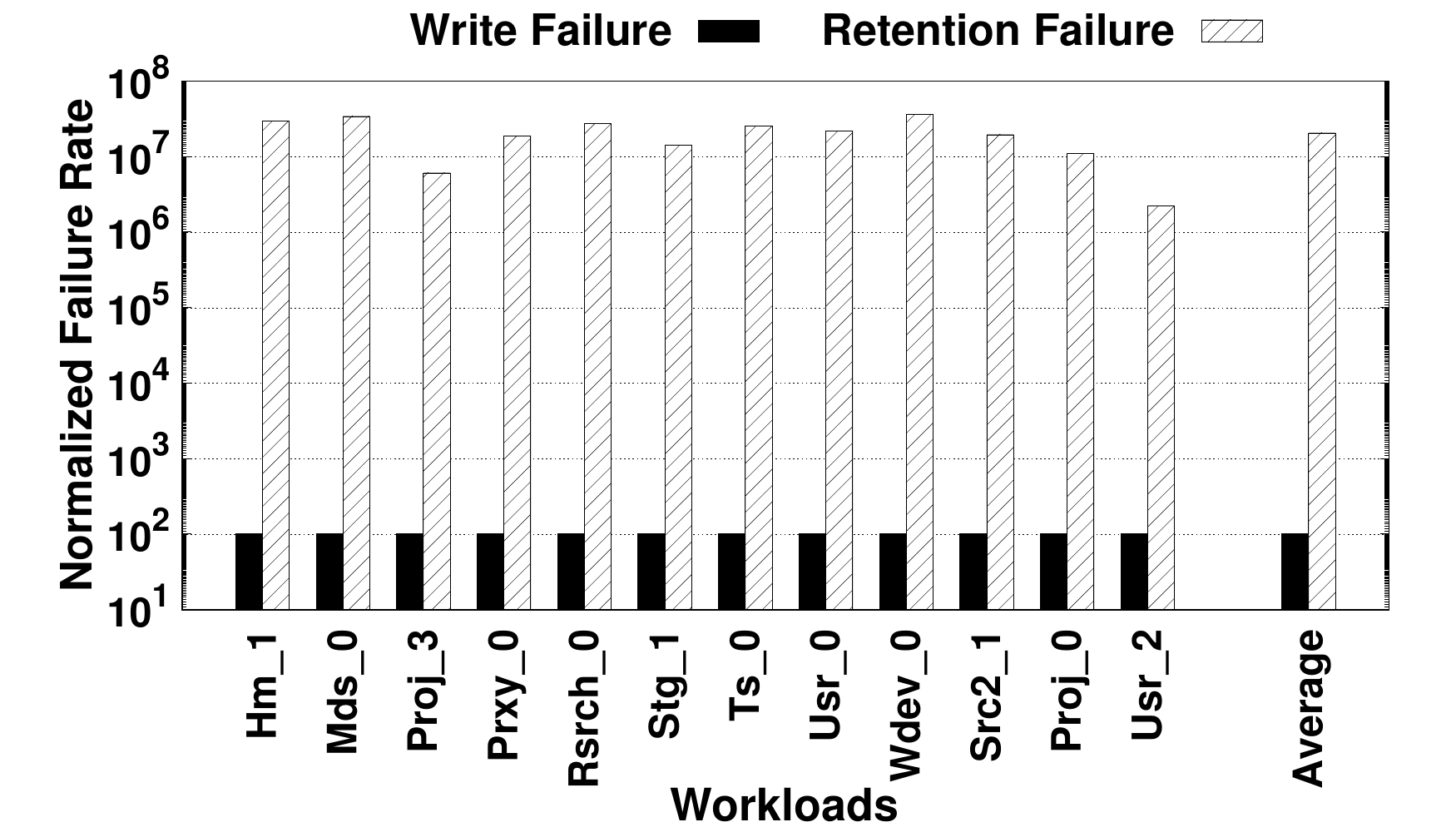}
	\centering\caption{Impact of each STT-MRAM error type on PJA failures}
	\label{Motive-FailureFootprint}
\end{figure} 

To investigate the probability of PJA data loss due to each STT-MRAM error type in recent technology nodes, we assume the probability of an erroneous write is $10^{-8}$ for an STT-MRAM cell \cite{DATECheshmikhani,TA-LRW,ITJ}. The probability of data loss for each error type is calculated based on equations (1)-(5). Note that read disturbance is not a major concern in PJA as it is only probable in case of data recovery where each PJA page is read once. 

\begin{figure}[t]
	\includegraphics[width=9cm]{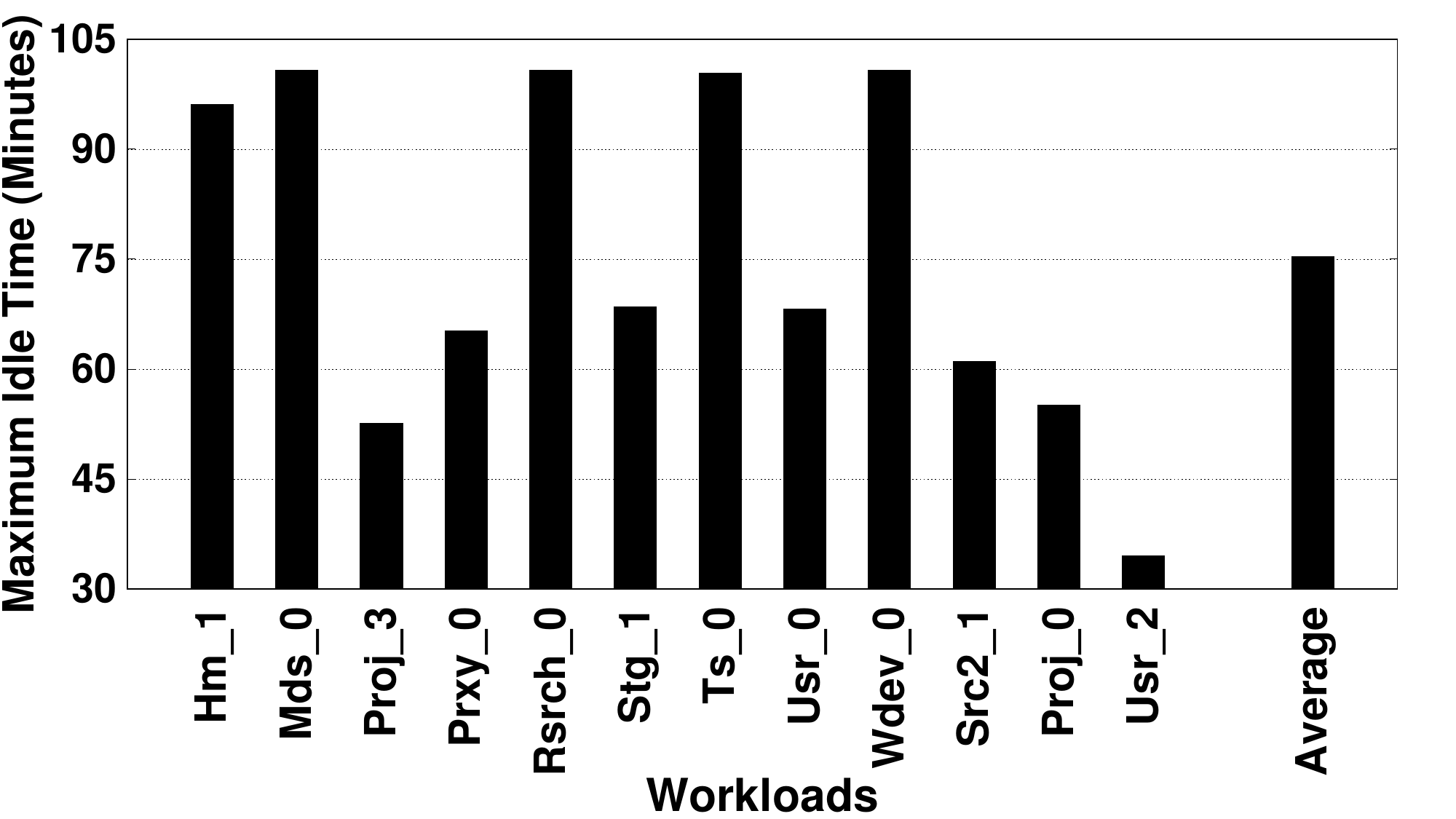}
	\centering\caption{Maximum page idle time of PJA}
	\label{Motive-Observ}
\end{figure} 
\begin{figure*}[t]
	\includegraphics[width=14.5cm]{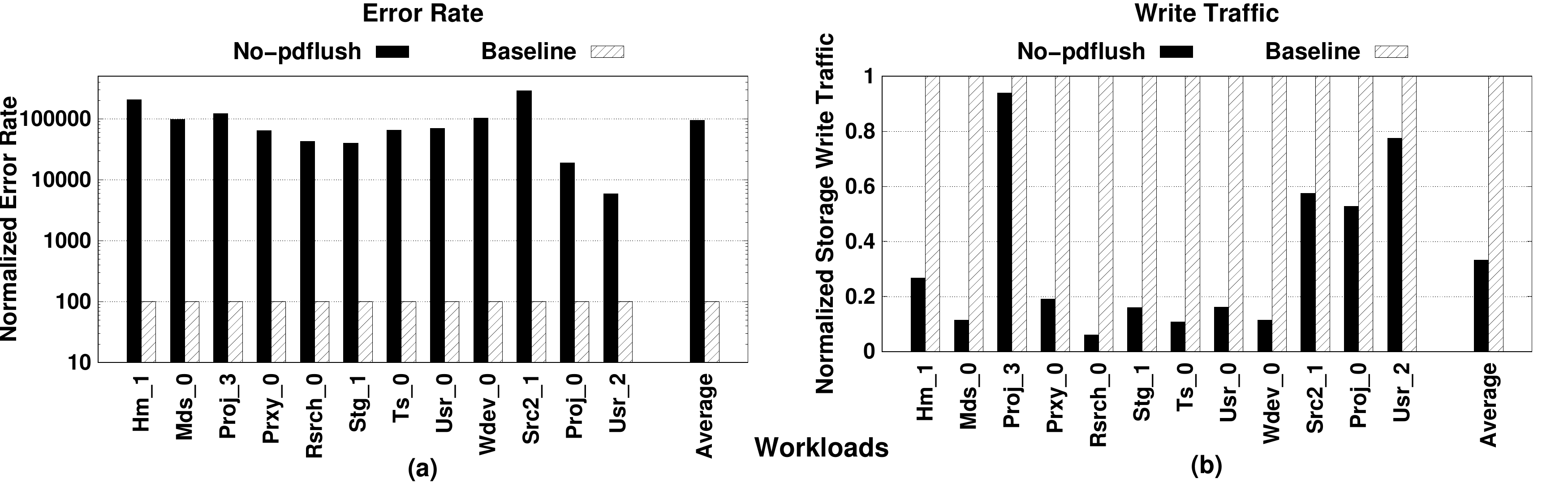}
	\centering\caption{Impact of periodic flushing on a) error rate and b) storage write traffic: No-pdflush (eliminated periodic flush) vs. Baseline}
	\label{Motive-PerBC-Failure}
\end{figure*}

Fig. \ref{Motive-FailureFootprint} shows the probability of PJA data loss, normalized to data loss due to write failure. Retention failure is the dominant source of failure as its footprint is drastically higher than write failure (an average of five orders of magnitude). Besides technology node characteristics, long and undetermined idle time of PJA pages is one of the main contributors to the data failure due to retention failure. Fig. \ref{Motive-Observ} shows the maximum page idle time of PJA pages. Neglecting page idleness as one of the design parameters of NVB-Buffer management scheme leads to pages with maximum idle time for an average of 75 minutes (up to 100.8 minutes), which makes them more vulnerable to retention failure. 

\vspace{-1pt}
One of the conventional approaches that can reduce page idle time in PJA is periodic flush, the same approach as Linux \textit{pdflush} function. In this approach, the dirty pages of NVB-Buffer are flushed to the main storage, based on a predetermined time period. Using periodic flush, an upper bound is set for maximum idle time of PJA pages as data is written to the main storage and PJA data pages can be discarded. Despite reducing page idle time, employing periodic flush considerably increases the storage write traffic \cite{DF-LRW}.

Fig. \ref{Motive-PerBC-Failure} shows the impact of \textit{using} (Baseline) and \textit{eliminating} periodic flush (No-pdflush) on PJA reliability and storage write traffic. In the baseline, dirty pages of NVB-Buffer are monitored and the flushing procedure is invoked every five seconds to flush the pages with 30-seconds idle time. 
Enabling periodic flush in the baseline reduces the failure rate by three orders of magnitude (940$\times$, on average) compared to No-pdflush by preventing long page idle time and guaranteeing an upper bound for data pages (depicted in Fig. \ref{Motive-PerBC-Failure}-a). Nevertheless, employing the baseline significantly increases storage write traffic (as shown in Fig. \ref{Motive-PerBC-Failure}-b). Although using No-pdflush increases the probability of data loss, it reduces storage write traffic by an average of 66.7\% (up to 93.9\% in Rsrch\_0), as it provides more chance of re-accessing for dirty data pages in NVB-Buffer. 

Designing an efficient NVB-Buffer needs approaches that address long idle time of PJA pages with respect to the storage write traffic. However, \textit{none} of the prior studies address the reliability challenges of STT-MRAM based PJA. Moreover, existing schemes for mitigating retention failure at circuit- and/or architecture-level are either expensive or inefficient and optimized for processor cache or main memory \cite{Sanitizer,Circuit}. Based on these observations, we investigate that there is a need for a system-level scheme to enhance the reliability of STT-MRAM based NVB-Buffers, which is the motivation of this work.
\section{Proposed Scheme}
\begin{figure}[b]
	\includegraphics[width=8cm]{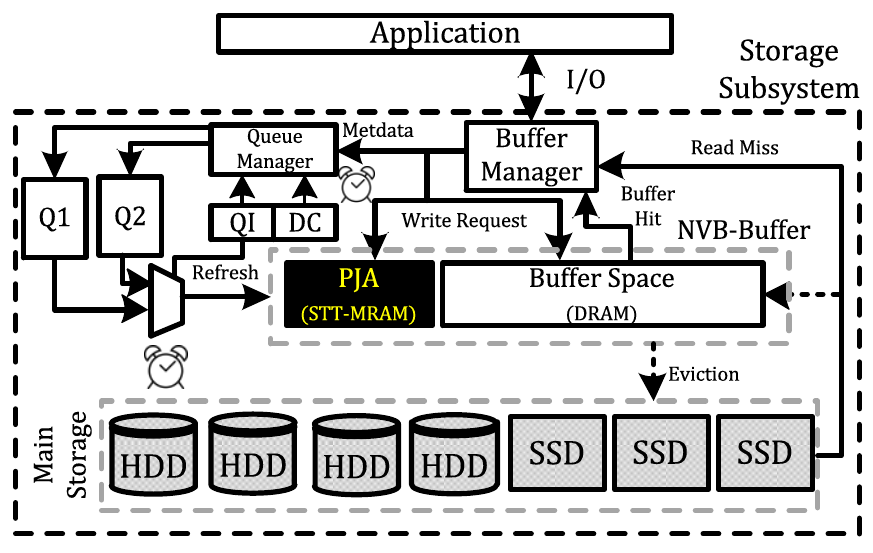}
	\centering\caption{Overview of CoPA}
	\label{Overview}
\end{figure}

A commercialized STT-MRAM based NVB-Buffer needs to reconcile its retention failure challenge. 
To this end, we propose a novel NVB-Buffer management scheme named \textit{\underline{Co}ld \underline{P}age \underline{A}wakening} (CoPA) to reduce the idle time of PJA pages. Fig. \ref{Overview} shows the overall structure of CoPA. While the buffer regularly performs its conventional functions such as a) admission of read/write requests to NVB-Buffer, b) handling buffer hits, and c) metadata management, it also aims at preventing retention failure using NVB-Buffer characteristics. CoPA employs \textit{Distant Refreshing} instead of conventional refreshing \cite{RetentionHPCA11} mechanism to overcome the large idle time of PJA pages. Regarding the features of recent technologies and NVB-Buffer structure, Distant Refreshing aims at refreshing PJA pages based on their corresponding replica in buffer space without increasing the probability of read disturbance. 

\vspace{-2pt}
CoPA controls the refreshing procedure by tracking the metadata of PJA pages using two queues and a simple 2-bits counter (bits are indicated as QI and DC in Fig. \ref{Overview}) to differentiate between recently-written and idle pages. Consequently, CoPA reduces the probability of retention failure by time-based refreshing of idle pages, without intensifying the probability of read disturbance. The rest of this section details Distant Refreshing mechanism and its benefits over conventional refreshing (Section 4.1), then provides a step-by-step depiction of page awakening procedure (Section 4.2), and finally presents the flow of CoPA for each request (Section 4.3). 


\begin{figure}[t]
	\includegraphics[width=9cm]{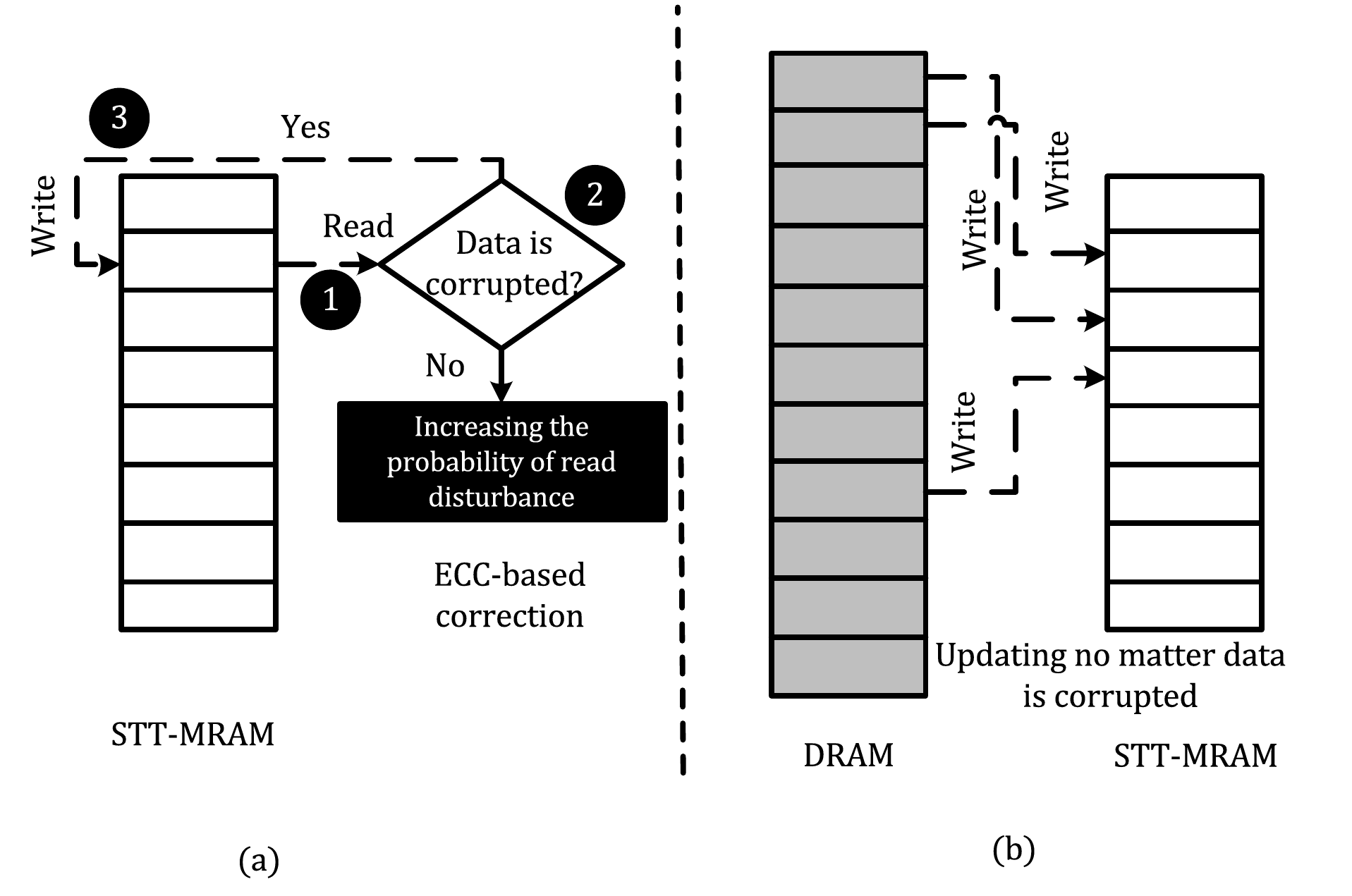}
	\centering\caption{Refresh approaches: a) conventional refreshing, and b) distant refreshing}
	\label{Update-Approaches}
\end{figure}

\begin{figure}[b]
	\includegraphics[width=9cm,height=6cm]{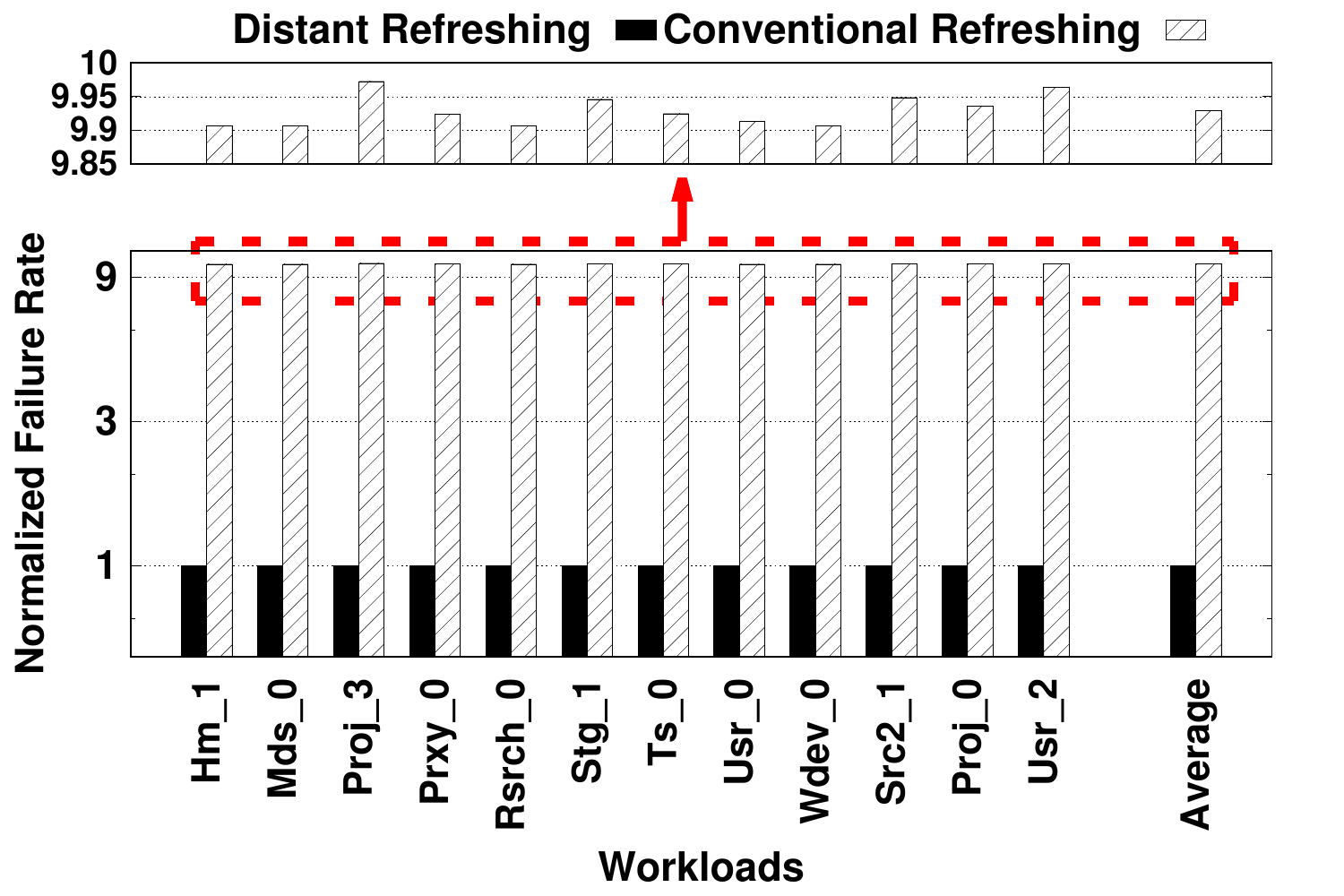}
	\centering\caption{Impact of distant refreshing vs. conventional refreshing on failure rate}
	\label{DistantImpact}
\end{figure}
\begin{figure*}[t]
	\includegraphics[width=14cm]{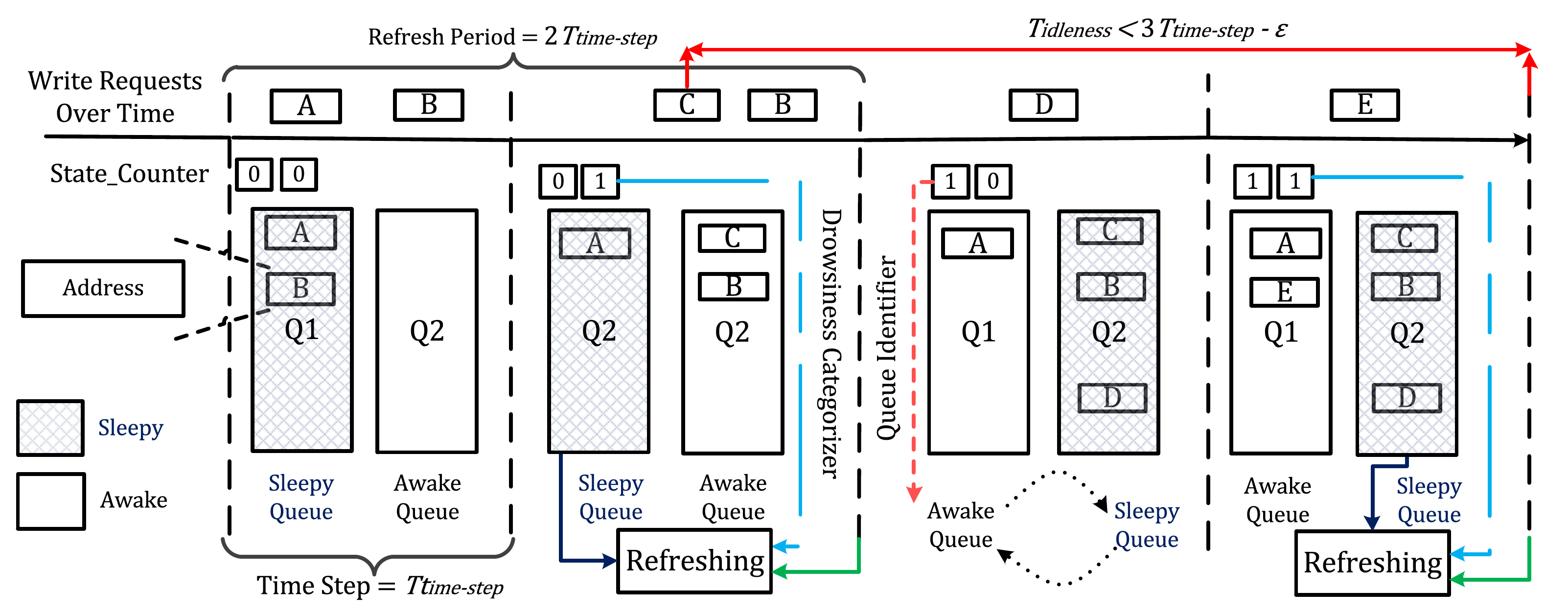}
	\centering\caption{Queue management in CoPA}
	\label{Counter_State}
\end{figure*}

\subsection{Page Awakening Procedure}
\subsubsection{Refreshing Aggression}
NVB-Buffer can provide a valid replica of each PJA page, which is necessary for Distant Refreshing. However, employing Distant Refreshing faces several challenges such as the length of refreshing intervals (\textit{Refresh Period}) or the way to select some pages for refreshing. In an aggressive approach, all PJA pages are refreshed with a short refresh period. This approach suffers from performance overhead and higher probability of write failure (because of successive write operations). Short refresh period along with refreshing all of PJA pages lead to reading a considerable amount of data from DRAM and writing them to STT-MRAM. Moreover, re-writing pages every few seconds in the aggressive refreshing method can increase the total data failure probability, as it imposes more write failure. 
Aggressive refreshing also increases the temperature due to extensively reading from DRAM and writing to STT-MRAM, which increases the total probability of failure \cite{TA-LRW}. Therefore, aggressive refreshing is \textit{not} an efficient approach for NVB-Buffers due to not only its reliability side effects, but also imposing significant memory overhead.

In a conservative refresh approach, the refresh period takes longer, while specific pages are refreshed. However, a long refresh period increases the probability of retention failure. Moreover, detecting proper pages for refreshing is another challenge of the conservative approach. A simple way to detect the best candidate is continuously checking the access time of each PJA page, which imposes memory overhead. Another solution is to employ detection mechanism such as a) reading the page content and detecting failures by ECC checking, b) comparing the page content with its corresponding copy in DRAM, or c) prediction \cite{Sanitizer,RetentionHPCA11}. However, these solutions \textit{either} cause extra read operations from PJA \textit{or} are not accurate. Thus, conservative refreshing is inefficient due to high memory overhead and still probable retention failure. The ideal scheme should balance between these two extreme approaches. 

\subsubsection{CoPA Queue Management}
CoPA a) increases the length of refresh intervals, b) guarantees an upper bound for idle time of each page, and c) prevents the recently-written pages from refreshing to reduce the overheads. To this end, CoPA takes advantage of two queues, named Q1 and Q2, and interprets them as \textit{Sleepy Queue} and \textit{Awake Queue} to differentiate and manage idle and recent pages. Sleepy queue tracks pages with high idle times while Awake queue tracks the metadata of pages with low idle times. CoPA also uses a 2-bit counter called \textit{State\_Counter} to manage both queues, as shown in Fig. \ref{Counter_State}. State\_Counter is employed for differentiating and managing Q1 and Q2.

CoPA divides each refresh period into two time-steps where at the end of each time-step, the value of the State\_Counter is increased by one. The \textit{Most Significant Bit} (MSB) and \textit{Least Significant Bit} (LSB) of the 
State\_
Counter are recognized as QI (Queue Identifier) and DC (Drowsiness Categorizer). Fig. \ref{Counter_State} shows how CoPA manages these queues, based on the value of State\_Counter. QI is responsible for identifying the Sleepy and Awake queues. When QI is `0', Q1 is recognized as the Sleepy queue, otherwise, Q2 is interpreted as the Sleepy queue. DC manages the insertion of each page metadata. If DC is `0', the metadata of incoming page is inserted into the Sleepy queue. On the other hand, the metadata of incoming page is inserted into the Awake queue. The refresh operation is performed at the end of the refresh period (after two time-steps), which leads to QI transition (from `0' to `1' or from `1' to `0') and refreshing PJA pages based on the Sleepy queue. Thus, CoPA gets rid of the high overhead of tracking and checking the idle time of each page by employing a simple State\_Counter. Moreover, using DC, CoPA is able to prevent refreshing of recently written pages and using QI, CoPA just changes the label of its queues (Q1 and Q2), instead of copying their contents. 

Fig. \ref{Counter_State} shows an example of how CoPA manages the refreshing procedure. It is assumed that the initial value of the State\_Counter is 0. In the first time-step, DC is `0' so \textit{A} and \textit{B} as incoming requests are inserted into the Sleepy queue. As QI is `0', Q1 and Q2 are interpreted as the Sleepy and Awake queues, respectively. At the end of each time-step, the value of State\_Counter is increased by one. In the second time-step, as DC is `1', the metadata of requests are redirected to the Awake queue. Since QI value is not changed, Q1 and Q2 are still the Sleepy and Awake queues, respectively. The metadata of \textit{C} is inserted into Q2 and for \textit{B}, the old version in Q1 is invalidated and then the metadata of \textit{B} is inserted into Q2.

At the end of the second time-step, the refresh procedure is performed based on entries of the Sleepy queue, which is Q1. Thus, the corresponding page of \textit{A} in PJA is refreshed. By increasing State\_Counter value, the QI value is changed from `0' to `1'. Hence, the labels of Q1 and Q2 are interchanged, i.e., Q1 becomes Awake queue and Q2 is now the Sleepy queue. As DC is `0', the metadata of \textit{D} is inserted into the Sleepy queue (Q2). At the fourth time-step, based on DC value (`1'), \textit{E}'s metadata is inserted into the Awake queue (Q1). At the end of the fourth time-step, the refreshing of \textit{C}, \textit{B}, and \textit{D} is performed as their addresses reside in the Sleepy queue (Q2). At the worst case, each page is refreshed after three time-steps. Therefore, if a page is not re-written by the application, the maximum idle time of each page is according to (\ref{eq-CoPA}):

\vspace{-10pt}{\normalsize \begin{equation} \label{eq-CoPA}
		T_{time-step} + \epsilon < T_{idleness} < 3T_{time-step} - \epsilon,
\end{equation}}where \textit{T\textsubscript{timestep}} is the time-step period (the half of each refresh period), \textit{T\textsubscript{idleness}} is the page idleness time, and $\epsilon$ is the minimum time unit.

\begin{figure}[t]
	\includegraphics[width=9cm]{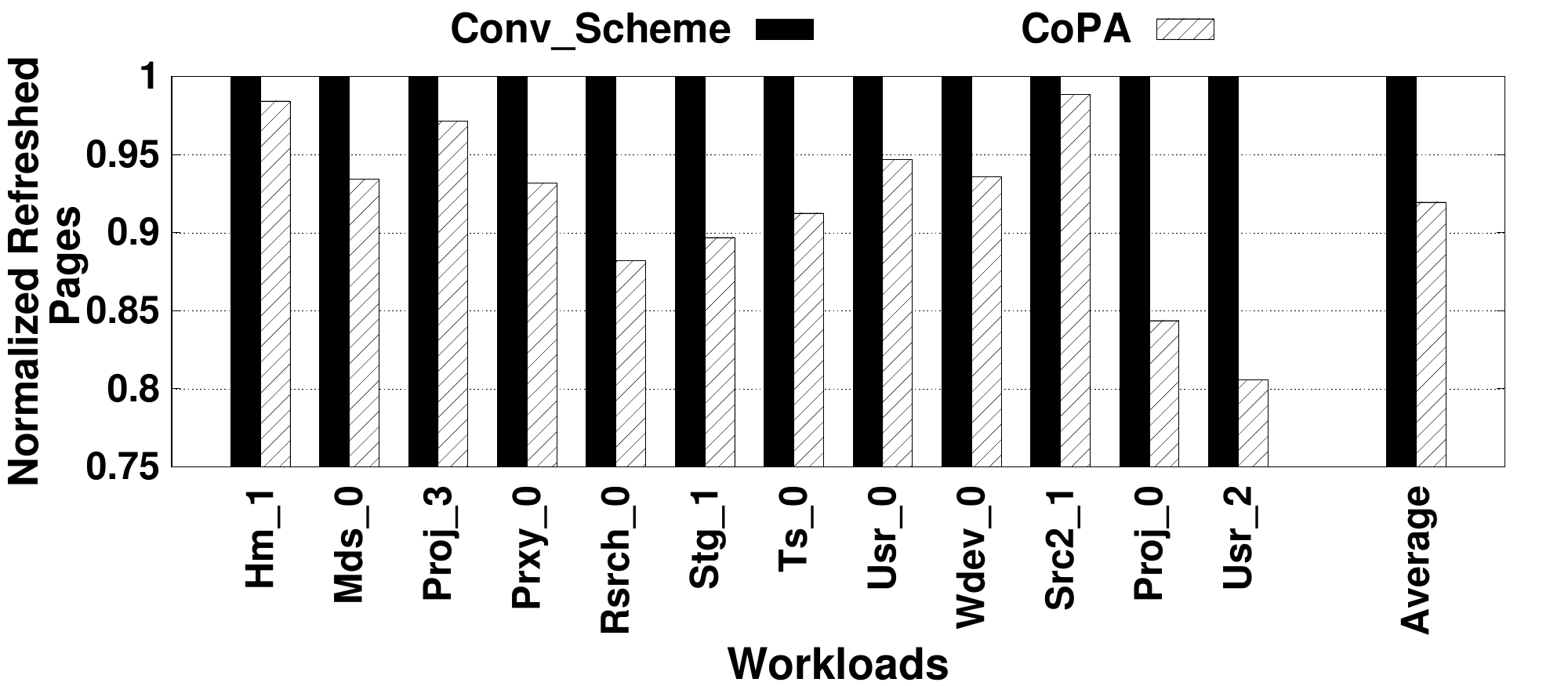}
	\centering\caption{Number of refreshed pages}
	\label{COPAEFF}
\end{figure}

We evaluate CoPA to compare with aggressive refreshing (Conv\_Scheme) as a conventional scheme that aims to refresh PJA pages by a predetermined period equal to 60 seconds. Fig. \ref{COPAEFF} shows the number of refreshed pages normalized to Conv\_Scheme. CoPA reduces the number of refreshed pages up to 19.5\% compared to Conv\_Scheme, as it prevents the refreshing of recently written pages. By reducing the refreshing period of Conv\_Scheme, the gap between CoPA and Conv\_Scheme is increased.

\begin{algorithm}[t]
	
	\small\begin{flushleft}
		
		{
			
			State\_Counter: 2-bits
			
			
			
			Q1, Q2: CoPA Queues \;
			
			\textbf{Def.} QI: MSB of State\_Counter \qquad //\textit{Queue Identifier}
			
			\textbf{Def.} DC: LSB of State\_Counter \hspace{0.5cm} //\textit{Drowsiness Categorizer}
			
			\textbf{Def.} Q1 is Sleepy\_Queue \& Q2 is Awake\_Queue if QI == 0
			
			\textbf{Def.}	Q2 is Sleepy\_Queue \& Q1 is Awake\_Queue if QI == 1
			
			\textbf{Initial} State\_Counter=0,  \;
			
			
			
			\
			
			\textbf{Procedure} PJA\_Refreshing
			
			\Begin
			{		
				
				\If{$DC = 1$}
				{
					// \textit{Refreshing Operation}
					
					
					Refresh PJA Pages, Based on Sleepy\_Queue
					

				}
				State\_Counter = State\_Counter + 1

			}
			
			\
			
			\textbf{Procedure} Req\_Management()
			
			\Begin
			{
				\For{Each new request \textit{P}}
				{
					Insert P in NVB\_Buffer
					
					\If{P is write request}
					{
						Invalid P.adrr If It Was Existed in Q1 or Q2
						
						
						
						\eIf{$DC = 0$}
						{
							//\textit{Sleepy\_Queue insertion}
							
							Insert P.addr in Sleepy Queue
							
							
						}
						{
							//\textit{Awake\_Queue Insertion}
							
							Insert P.addr P in Awake Queue
							
						}
					}
					
					\If{Dirty page \textit{E} is evicted from NVB-Buffer}
					{
						Invalid E.adrr If It Was Existed in Q1 or Q2
					}
				}
			}

		}
	\end{flushleft}
	\caption{Procedure of Page Awakening}
	\label{Algorithm-Proc-of-CoPA}	
\end{algorithm}

\subsection{Putting It All Together}
Algorithm \ref{Algorithm-Proc-of-CoPA} shows the flow of CoPA scheme. CoPA mainly consists of two procedures: 1) $Req\_Man$-
$agement$ and 2) $PJA\_$ $Refreshing$. \textit{State\_Counter} is a 2-bit counter, which is used to distinguish between the Sleepy queue and Awake queue, based on its MSB (QI) and LSB (DC) values. 
Each entry of these queues consists of the page address of the inserted page.

$PJA\_$ $Refreshing$ is invoked based on the time-steps by a timer interrupt. Based on the value of DC, CoPA aims to refresh PJA pages. In $PJA\_$ $Refreshing$, the current value of DC is checked to determine that which queue is responsible for tracking incoming requests, and if it is `1', it means that the recent write requests are inserted into the Awake queue and now it is the time for QI transition. Hence, CoPA refreshes PJA pages based on the addresses of the Sleepy queue. 
Every time $PJA\_$ $Refreshing$ is invoked, the value of \textit{State\_Counter} is increased by one, which specifies the state of CoPA queues for the next time-step.

\begin{figure}[b]
	\includegraphics[width=9.5cm]{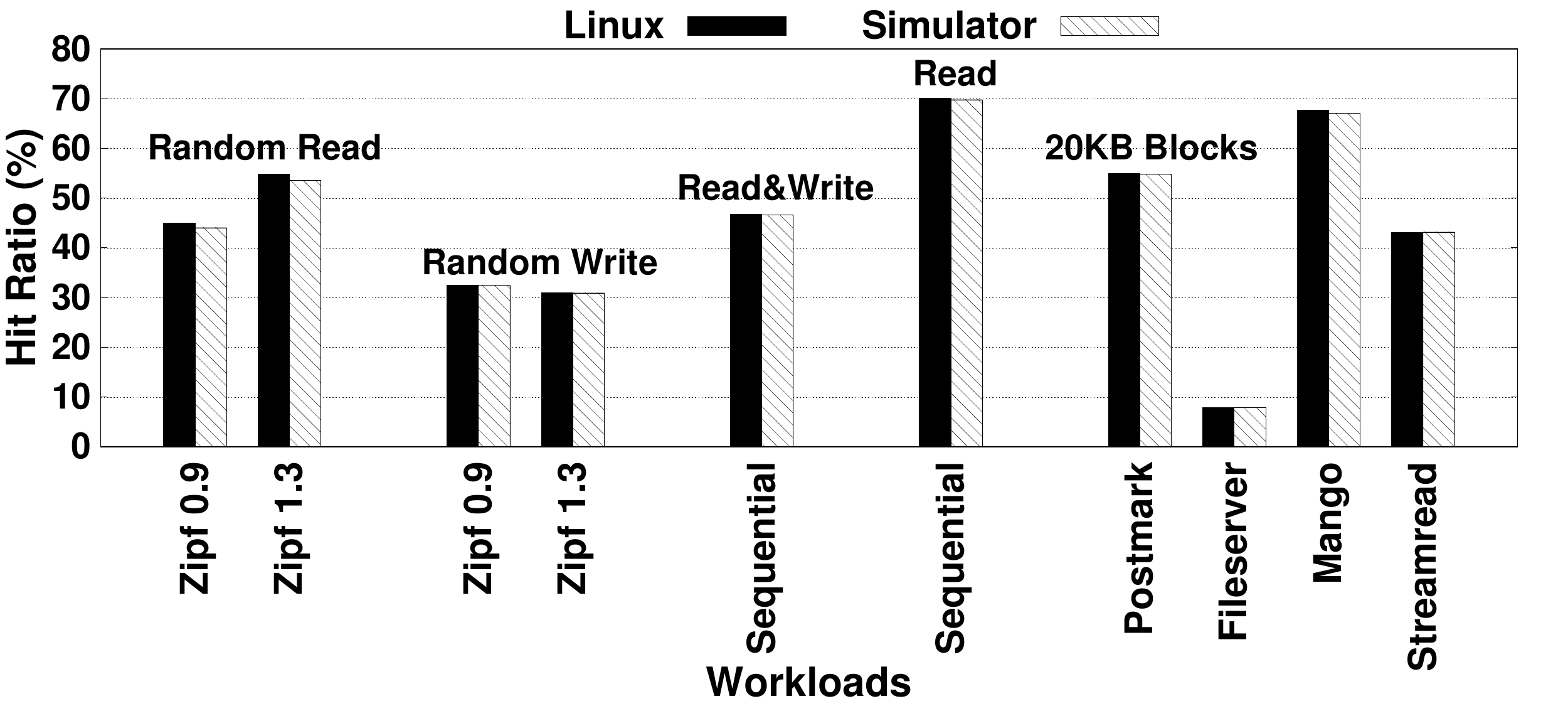}
	\centering\caption{Accuracy of the buffer cache simulator used for evaluations}
	\label{Simulat}
\end{figure}

CoPA also updates its queues by the incoming new write requests. If the page already exists in either of the CoPA queues, the corresponding page is invalidated. Based on the value of QI and DC, CoPA inserts the metadata of this page into the proper queue (into the Sleepy queue if DC is '0' and into the Awake queue, otherwise). Moreover, if a dirty page is evicted from NVB-Buffer, CoPA invalidates its metadata in CoPA queues. 

\begin{figure*}[t]
	\includegraphics[width=16cm]{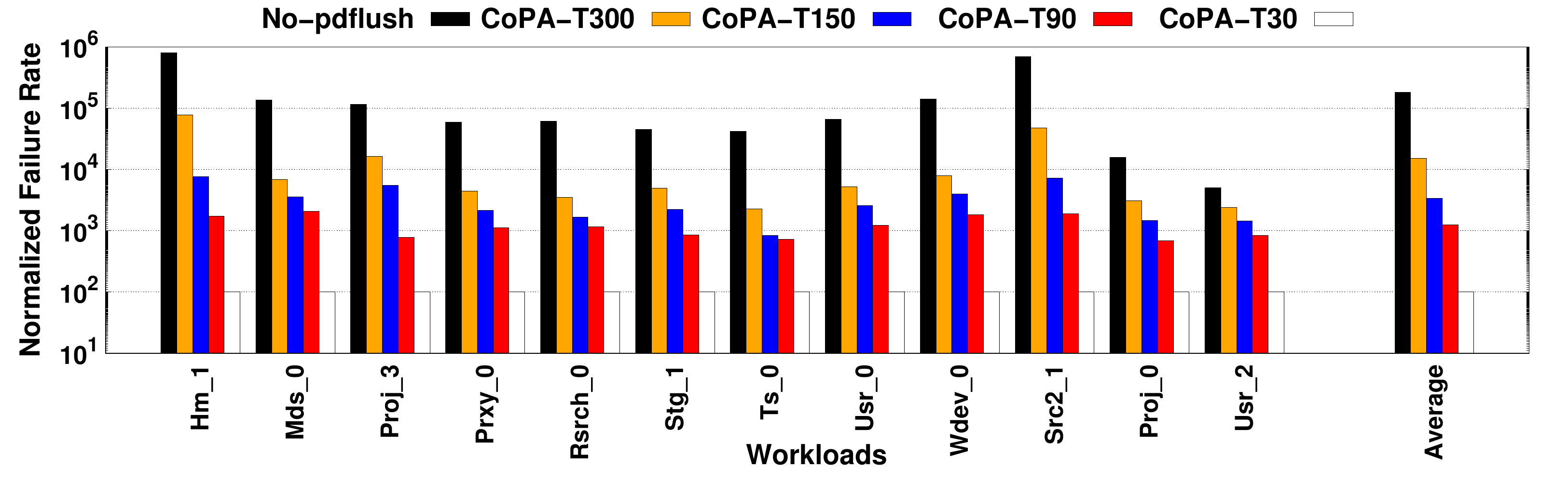}
	\centering\caption{Failure rate of CoPA (with different time-steps) compared to No-pdflush}
	\label{Result-Max-Idle}
\end{figure*}
\begin{figure*}[t]
	\includegraphics[width=14cm]{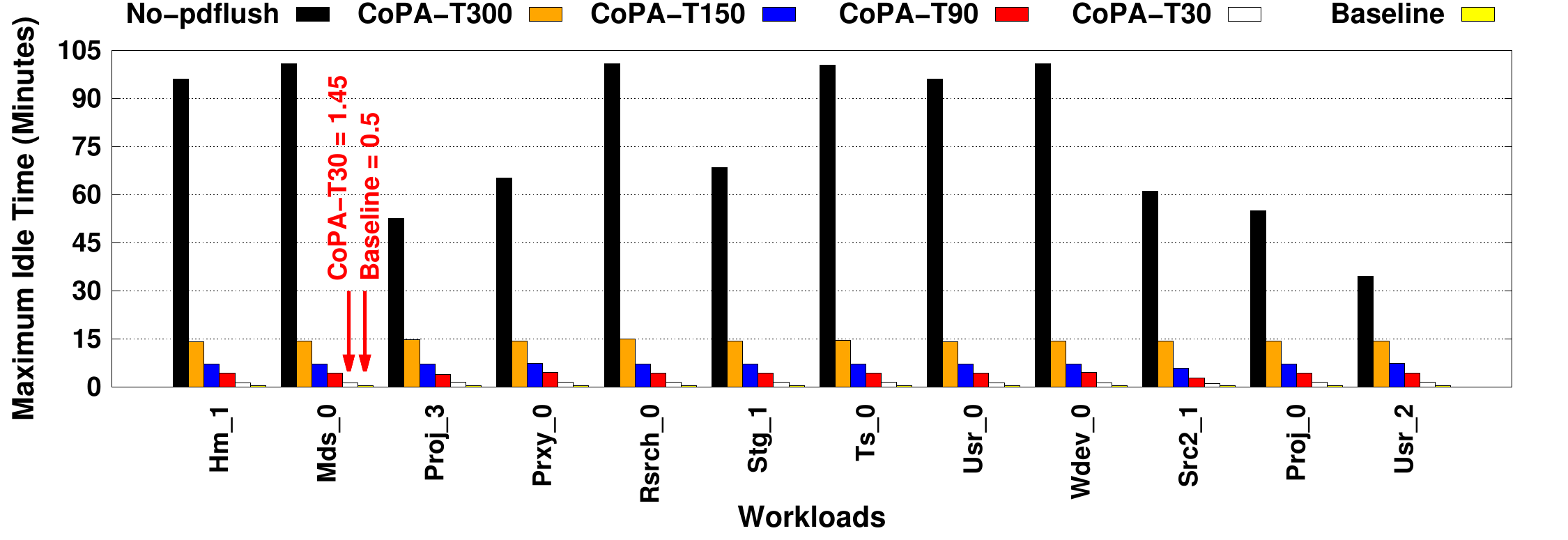}
	\centering\caption{Maximum idle time of CoPA compared to No-pdflush. Maximum idle time values: \textit{No-pdflush} (100.77 minutes), \textit{CoPA-T300} (14.95 minutes), \textit{CoPA-T150} (7.47 minutes), \textit{CoPA-T90} (4.5 minutes), and \textit{CoPA-T30} (1.5 minutes).}
	\label{Max-Idle}
\end{figure*}

\section{Evaluation}

\subsection{Evaluation Flow}
To evaluate CoPA, we take advantage of our in-house buffer cache simulator, which is developed based on  Linux kernel 4.4 and EXT2 filesystem, validated using Flexible I/O (fio) \cite{fio}, Filebench \cite{filebench} and Postmark \cite{Postmark} over different types of workloads. As shown in Fig. \ref{Simulat}, there is a negligible difference (up to 2.3\%) between the simulator and Linux hit ratio under different patterns of accesses (random or sequential) and various distributions (e.g., zipf 0.9 and zipf 1.3), which is due to the difference between employed LRU and the semi-LRU used by Linux. We extend the simulator for: 1) analyzing timing parameters of the application block requests and 2) investigating the impact of each scheme on the response time.  
The simulator is also equipped with DRAMSim2 \cite{DRAMSim2}, configured for simulating two DDR3 modules based on DRAM and STT-MRAM characteristics for the buffer and PJA, respectively \cite{STTCharac}.
We post-process the simulator outputs based on (1)-(5) to evaluate each scheme in term of retention failure probability.


Our evaluations are performed on twelve workloads from \textit{Microsoft Research Cambridge} (MSRC) traces \cite{MSRC}, based on an NVB-Buffer consists of 8GB DRAM as buffer space, 512MB STT-MRAM as PJA, and the page size of 4KB. Each of these pages are consisted of 512 64-bit datawords, each of which is protected by SEC-DED(64,72) code. CoPA is compared with the the state-of-the-art approaches \cite{DF-LRW,Unified-NVM-DRAM,Zhang-ICCD}, where the periodic flushing is eliminated, referred to as \textit{No-pdflush} (the physical journaling is considered as the journaling mechanism). To provide an overview of CoPA efficiency in term of performance, we also compare CoPA with a conventional reliable scheme as a \textit{Baseline}. The Baseline uses periodic flush based on 5-seconds intervals, where pages with 30 seconds idle time are flushed to the main storage and discarded from PJA, which is the default configuration in operating systems such as Linux \cite{DF-LRW}. Thus, this scheme reduces the probability of retention failures by providing a small idle time for PJA pages, but imposes significant storage write traffic compared to No-pdflush (as shown in Section 3.3.2).
\subsection{Failure Rate Analysis}

To investigate CoPA in term of reliability, we evaluate it based on different refresh periods and set the time-steps equal to 30 seconds (CoPA-T30), 90 seconds (CoPA-T90), 150 seconds (CoPA-T150), and 300 seconds (CoPA-T300). Fig. \ref{Result-Max-Idle} shows the failure rate of CoPA and No-pdflush normalized to CoPA-T30. 
On average, CoPA reduces the probability of failure by three orders of magnitude compared to No-pdflush. 

CoPA provides a significant idle time reduction for PJA pages (up to 66.9$\times$) by guaranteeing to refresh each PJA page at least every three time-steps, as shown in Fig. \ref{Max-Idle}. 
It also provides a predictable PJA idle time to fill the gap between No-pdflush and the baseline approach in term of reliability, as there is no upper bound for PJA page idle time in No-pdflush scheme. There is a considerable difference between the maximum page idle time of different workloads (more than 66 minutes) in No-pdflush, while this value in CoPA differs in few seconds. Therefore, CoPA alleviates the variation of page idleness compared to the schemes that have eliminated periodic flush, and accordingly, provides a reliable scheme without increasing main storage traffic. 
CoPA can be tuned to provide the same reliability as the Baseline. However, in this case, the trade-off between the number of refreshed pages and failure rate period should be considered, as lower refreshing period leads to higher number of refresh operations. 


\subsection{Response Time}

\begin{figure*}[t]
	\includegraphics[width=14.5cm]{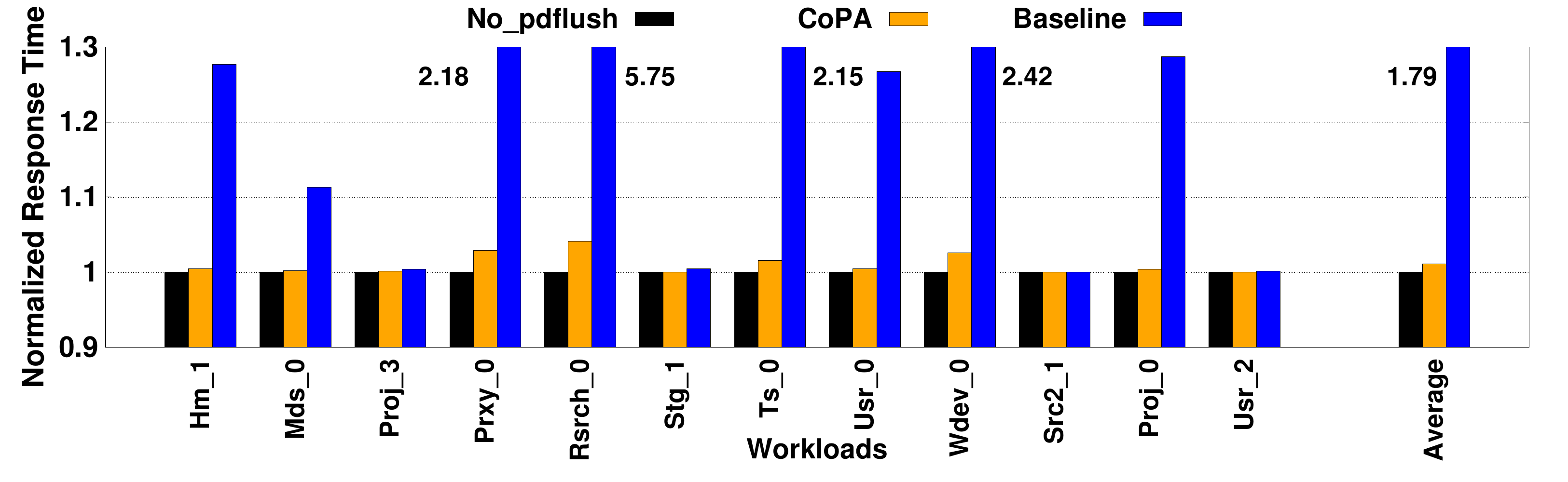}
	\centering\caption{Normalized response time of No\_pdflush, CoPA, and baseline}
	\label{Result-Storage-Traffic}
\end{figure*}

One of the key aspects of an efficient NVB-Buffer management scheme is its impact on the response time of storage subsystem. Although the baseline considerably reduces the probability of retention failure, it significantly increases storage write traffic compared to No-pdflush. However, CoPA not only significantly reduces the probability of retention failures compared to No-pdflush, but also avoids the baseline high storage write traffic. CoPA, as well as improving reliability compared to No-pdflush, aims at preventing the baseline poor write traffic. 

Fig. \ref{Result-Storage-Traffic} shows the response time of each scheme normalized to No-pdflush. The time-step of CoPA is set to 30 seconds, which brings the highest response time (due to higher refreshing operations). 
CoPA increases response time by an average of 1.1\% compared to No\_pdflush (up to 4.2\%), which is due to extra read and write operations on PJA and buffer. On the other hand, CoPA significantly reduces the response time compared to the Baseline. The reliability improvement provided by the Baseline is achieved by enabling periodic flushing, which leads to a considerable increase in storage write traffic. CoPA, however, enhances reliability without imposing extra storage write traffic, therefor it can offer a reliable NVB-Buffer management scheme with higher performance compared to the Baseline. 


System designers have the opportunity to get the performance behavior of CoPA closer to No-pdflush, by tuning time-step knob. With proper refresh period, CoPA can provide response time close to No-pdflush without long idle time for PJA pages. However, as already discussed, increasing time-step leads to higher failure rate. 
Nevertheless, it still can provide an upper bound for PJA pages, and yet higher reliability compared to No\_pdflush.

\subsection{Overall}

\section{Related Work \& Discussion}
Employing NVMs in buffering schemes has already been addressed in prior studies based on two main approaches: 1) storage-aware approach \cite{H-ARC,CLOCK-DNV,CFLRU} and 2) storage-unaware approach \cite{DF-LRW,Unified-NVM-DRAM,SN,REWIND,TWOLRU,NOVA,SCHEN-TCAD18,UniBuffer}. 
The first approach aims at improving the lifetime and performance of SSD-based storage systems using different mechanisms such as dirty page favoring \cite{H-ARC} (the mechanism of using policies to prioritize dirty pages in the buffer), overcoming small write problem, and reducing garbage collection frequency \cite{DEFT}. These schemes, however, increase the idle time of dirty pages in the NVM, which increase the probability of retention failure.

Schemes in the second approach employ NVMs as part of the main memory \cite{REWIND,NOVA,TWOLRU} or log/journal area \cite{DF-LRW,Unified-NVM-DRAM,SCHEN-TCAD18,SN}. As conventional schemes exploit a part of the main storage (HDD or SSD) as the journal area for crash recovery, this approach takes advantage of NVM persistency to reduce storage traffic by committing journaled data to the NVM. Nevertheless, these schemes are optimized toward performance improvement by decreasing storage write traffic, which is achieved by increasing the residency time of dirty pages in the NVM. Therefore, the schemes in the second approach increase the probability of retention failure compared to the conventional schemes, in favor of reducing storage write traffic. 

In \cite{mwJFS}, mwJFS is proposed to address the retention time of \textit{Multi-Level Cell} (MLC) PCM-based  jouranling file systems by employing different write pulse widths (and accordingly retention time). Although expanding the write pulse in order to increase the retention time is a conventional solution in NVMs, and it is also applicable to STT-MRAM, it leads to considerable increase in write latency and power consumption \cite{CacheRevive}. Moreover, employing the retention monitoring mechanism of mwJFS for STT-MRAM based PJAs increases the probability of read disturbance. CoPA, however, significantly decreases the error rate by reducing the page idle time compared to these schemes without increasing read disturbance probability, while providing a noticeable reduction of storage write traffic compared to the conventional schemes. 

CoPA is also applicable to the main memory buffer cache layer. To this end, \textit{Operating System} (OS) needs to assign a small portion of the main memory to the \textit{Awake} and \textit{Sleepy} queues. For example, for an 8GB main memory and 512MB PJA, OS should assign 1.2\% of the main memory space to CoPA queues. To employ CoPA for systems with logical journaling, a portion of DRAM equal to the size of NVM should be considered to store the required valid copy of journaled data for Distant Refreshing. Logical journaling needs much less space since the updates to the data pages are buffered. Thus, the required DRAM space for Distant Refreshing compared to the total DRAM space is negligible.

\section{Conclusion}
Employing STT-MRAM as PJA provides the opportunity for designing efficient NVB-Buffer schemes. The technology downscaling increases the probability of retention failure in STT-MRAM, especially in recent technologies where retention failure becomes the main source of errors. Although existing NVB-Buffer management schemes provide considerable storage write traffic reduction compared to the conventional schemes, they suffer from long and undetermined PJA page idle time. The longer the page idle time, the more the error rate compared to the conventional schemes.  
This paper proposed an efficient scheme named CoPA, for reducing the probability of retention failure. CoPA utilized an NVB-Buffer-friendly approach named \textit{Distant Refreshing}, which re-writes the idle PJA pages based on their replica in DRAM no matter it is corrupted or not. CoPA  monitors PJA pages using two queues (Awake and Sleepy queues) to distinguish pages with long idle time. 
CoPA also guarantees a tunable upper bound for maximum idle time of PJA pages. Our evaluations illustrated that CoPA reduces the probability of failure by three orders of magnitude with negligible performance overhead.

\bibliographystyle{IEEEtran}

\begin{IEEEbiography}[{\includegraphics[width=1in,height=1.25in,clip,keepaspectratio]{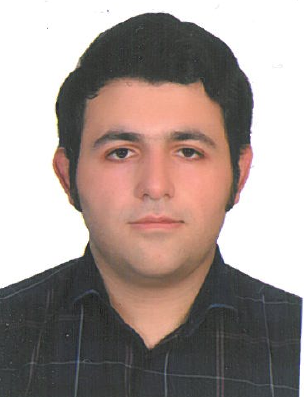}}]{\textbf{Mostafa Hadizadeh}}
	received the B.Sc. degree in computer engineering from Shahid Beheshti University (SBU), Tehran, Iran, in 2016, and the M.Sc. degree in computer engineering at Sharif University of Technology (SUT), Tehran, Iran, in 2018. He is a member of Data Storage, Networks, and Processing (DSN) Laboratory since 2017. From December 2016 to May 2017, he was a member of Dependable Systems Laboratory (DSL) at SUT. His research interests include computer architecture, memory systems, dependable systems, and systems on chip.
\end{IEEEbiography}
\vskip -3.5\baselineskip plus -1fil
\vfill
\begin{IEEEbiography}[{\includegraphics[width=1in,height=1.25in,clip,keepaspectratio]{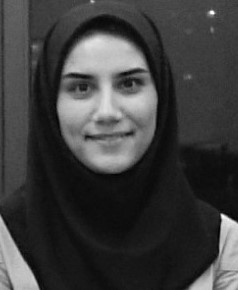}}]{\textbf{Elham Cheshmikhani}}
	received the B.Sc. degree in computer engineering from Iran University of Science and Technology (IUST), the M.Sc. degree in computer engineering from Amirkabir University of Technology (Tehran Polytechnic), Tehran, Iran, in 2011 and 2013, respectively and the PhD degree in computer engineering from Sharif University of Technology (SUT), Tehran, Iran in Feb. 2020. She was a member of the Design and Analysis of Dependable Systems (DADS) at AUT from 2011 to 2015 and has been a member of the Dependable Systems Laboratory (DSL) and Data
	Storage, Networks \& Processing Laboratory (DSN) since 2015 and 2017, respectively. Her research interests include emerging nonvolatile memory technologies, dependability analysis, fault tolerance, and storage systems.
	More recently, she received the Best Paper Award at IEEE/ACM Design, Automation, and Test in Europe (DATE) in 2019.
\end{IEEEbiography}
\vskip -3.5\baselineskip plus -1fil
\vfill
\begin{IEEEbiography}[{\includegraphics[width=1in,height=1.25in,clip,keepaspectratio]{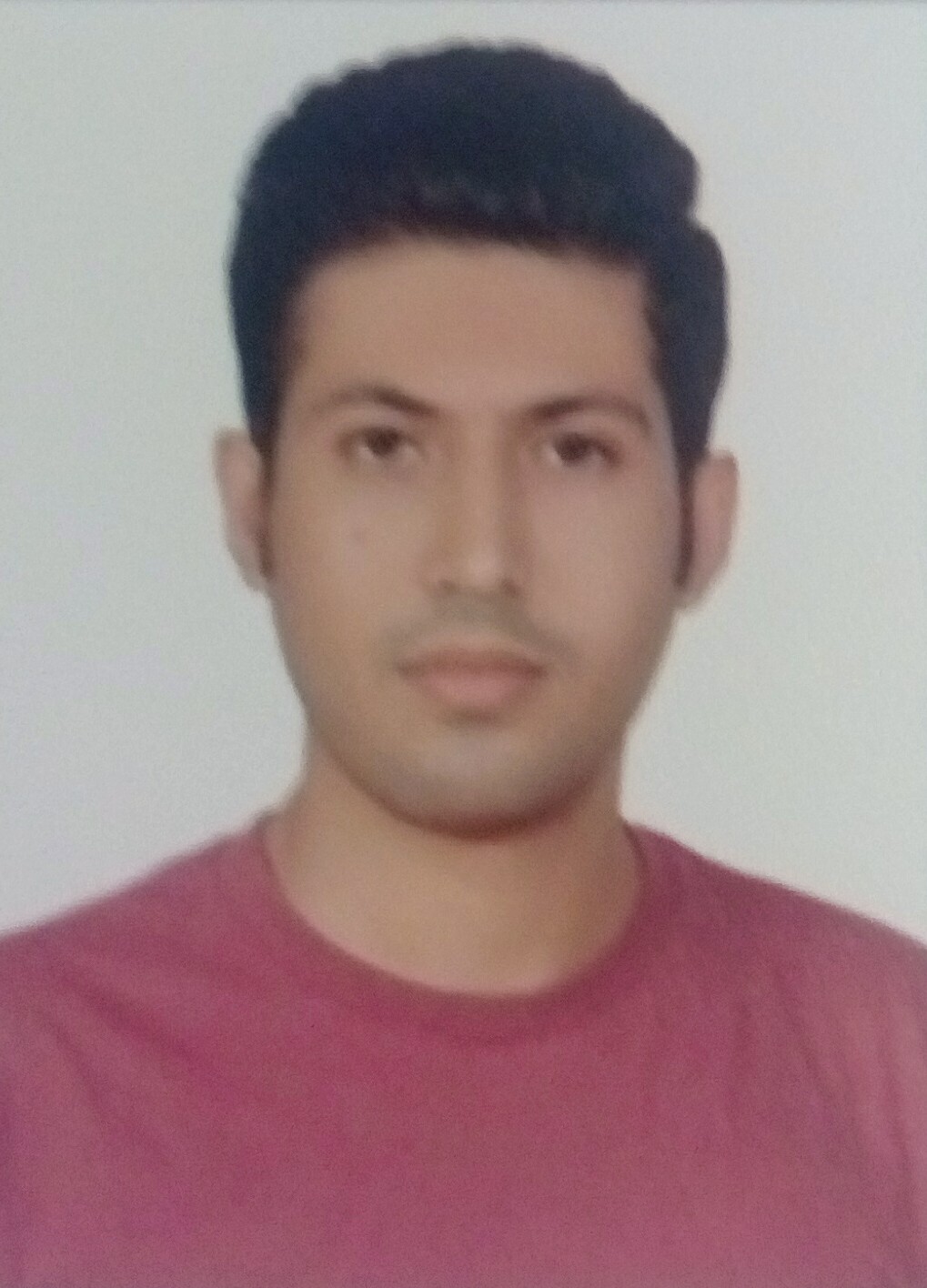}}]{\textbf{Maysam Rahmanpour}}
	received a B.Sc. degree in computer engineering from Shahid Beheshti University (SBU), Tehran, Iran in 2016. He received an M.Sc. degree in computer engineering from Sharif University of Technology (SUT), Tehran, Iran. He is a member of the Data Storage, Networks, and Processing (DSN) Laboratory since December 2016. His research interest includes Computer Architecture, Emerging NVM-Based Architecture, Memory Systems, and High-Performance Systems.
\end{IEEEbiography}
\vskip -3.5\baselineskip plus -1fil
\vfill
\begin{IEEEbiography}[{\includegraphics[width=1in,height=1.25in,clip,keepaspectratio]{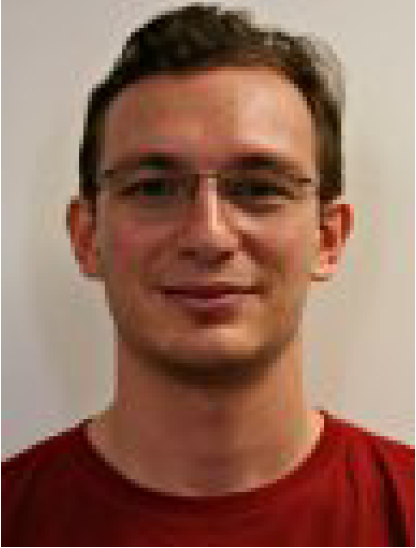}}]{\textbf{Onur Mutlu}}
	is a Professor of Computer Science at ETH Zurich. He is also a faculty member at
	Carnegie Mellon University, where he previously held the Strecker Early Career Professorship. His current broader research interests are in computer architecture, systems, hardware security, and bioinformatics. A variety of techniques he, along with his group and collaborators, has invented over the years have influenced industry and have been employed in commercial microprocessors
	and memory/storage systems. He obtained his PhD and MS in ECE from the University
	of Texas at Austin and BS degrees in Computer Engineering and Psychology from the University of Michigan, Ann Arbor. He started the Computer Architecture Group at Microsoft Research (2006-2009), and held various product and research positions at Intel Corporation, Advanced Micro Devices, VMware, and Google. He received the inaugural IEEE Computer Society Young Computer Architect Award, the inaugural
	Intel Early Career Faculty Award, US National Science Foundation CAREER Award, Carnegie Mellon University Ladd Research Award, faculty partnership awards from various companies, and a healthy number of best paper or ”Top Pick” paper recognitions at various computer systems, architecture, and hardware security venues. He is an ACM Fellow ``for contributions to computer architecture research, especially in memory systems'', IEEE Fellow for ”contributions to computer architecture research and practice”, and an elected member of the Academy of Europe
	(Academia Europaea). His computer architecture and digital circuit design course lectures and materials are freely available on YouTube, and his research group makes a wide variety of software and hardware artifacts freely available online. For more information, please see his webpage at https://people.inf.ethz.ch/omutlu/.
\end{IEEEbiography}
\vskip -3.5\baselineskip plus -1fil
\vfill
\begin{IEEEbiography}[{\includegraphics[width=1in,height=1.25in,clip,keepaspectratio]{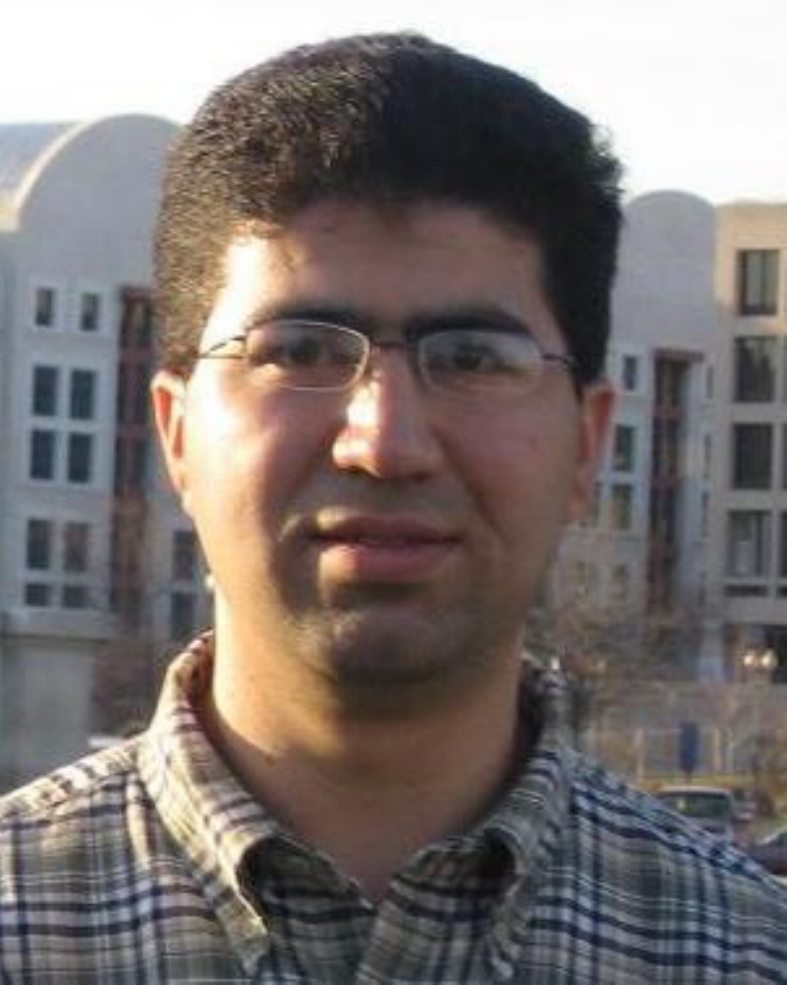}}]{\textbf{Hossein Asadi}}
	(M'08, SM'14) received the BSc and MSc degrees in computer engineering from the SUT, Tehran, Iran, in 2000 and 2002, respectively, and the PhD degree in computer engineering from Northeastern University, Boston, MA, USA, in 2007.
	
	He was with EMC Corporation, Hopkinton, MA, as a research scientist and senior hardware engineer, from 2006 to 2009. From 2002 to 2003, he was a member of the Dependable Systems Laboratory, SUT, where he researched hardware verification techniques. From 2001 to 2002, he was a member of the Sharif Rescue Robots Group. He has been with the Department of Computer Engineering, SUT, since 2009, where he is currently a full professor. He is the founder and director of the \emph{Data Storage, Networks, and Processing} (DSN) Laboratory and the director of Sharif \emph{High-Performance Computing} (HPC) Center. He spent three months in the summer 2015 as a Visiting Professor at the School of Computer and Communication Sciences at EPFL. He is also the co-founder of HPDS corp., designing and fabricating midrange and high-end data storage systems. He has authored and co-authored more than eighty technical papers in reputed journals and conference proceedings and holds several international patents. His current research interests include data storage systems and networks, solid-state drives, operating system support for I/O and memory management, and high-performance, reconfigurable, and dependable computing.
	
	Dr. Asadi was a recipient of the Technical Award for the Best Robot Design from the International RoboCup Rescue Competition, organized by AAAI and RoboCup, a recipient of Best Paper Award at the 15th CSI International Symposium on \emph{Computer Architecture \& Digital Systems} (CADS), the Distinguished Lecturer Award from SUT in 2010, the Distinguished Researcher Award and the Distinguished Research Institute Award
	from SUT in 2016, the Distinguished Technology Award from SUT in 2017, and the Distinguished Research Lab Award from SUT in 2019.
	He is also recipient of Extraordinary Ability in Science visa from US Citizenship and Immigration Services in 2008. He has been ranked among ``Top-10'' among 500+ faculties by Research and Technology Deputy, Sharif University of Technology for five consecutive years from 2016 to 2020. More recently, he received the Best Paper Award at IEEE/ACM Design, Automation, and Test in Europe (DATE) in 2019. He has served as a guest editor of IEEE Transactions on Computers, an Associate Editor of Microelectronics Reliability, a Program Co-Chair of CADS2015, and the Program Chair of CSI National Computer Conference (CSICC2017). He is a senior member of the IEEE.
\end{IEEEbiography}

\end{document}